\input epsf
\documentclass[nofootinbib,aps,superscriptaddress,preprint,floatfix]{revtex4}

\usepackage{graphicx}

\def\D0bar{\overline D{}^0}
\def\K0bar{\overline K{}^0}
\def\DDbar{D^0 - \overline D{}^0}
\def\beq{\begin{equation}}
\def\eeq{\end{equation}}
\def\beqa{\begin{eqnarray}}
\def\eeqa{\end{eqnarray}}
\def\bea{\begin{eqnarray}}
\def\eea{\end{eqnarray}}
\def\beq{\begin{equation}}
\def\eeq{\end{equation}}

\def\Re{{\cal R \mskip-4mu \lower.1ex \hbox{\it e}\,}}
\def\Im{{\cal I \mskip-5mu \lower.1ex \hbox{\it m}\,}}
\def\be{\begin{equation}}
\def\ee{\end{equation}}

\def\Re{{\cal R \mskip-4mu \lower.1ex \hbox{\it e}\,}}
\def\Im{{\cal I \mskip-5mu \lower.1ex \hbox{\it m}\,}}

\def\sub#1{_{\lower.25ex\hbox{$\scriptstyle#1$}}}
\def\sul#1{_{\kern-.1em#1}}
\def\sll#1{_{\kern-.2em#1}}
\def\sbl#1{_{\kern-.1em\lower.25ex\hbox{$\scriptstyle#1$}}}
\def\ssb#1{_{\lower.25ex\hbox{$\scriptscriptstyle#1$}}}
\def\sbb#1{_{\lower.4ex\hbox{$\scriptstyle#1$}}}

\def\to{\rightarrow}
\def\dmix{\ifmmode D^0-\bar D^0 \else $D^0$-$\bar D^0$\fi}
\def\dm{\Delta M_D}

\def\dmd{\ifmmode \Delta M_D \else $\Delta M_D$\fi}
\def\mh{\ifmmode m\sbl H \else $m\sbl H$\fi}
\def\mch{\ifmmode M_{H^\pm} \else $M_{H^\pm}$\fi}
\def\mt{\ifmmode m_t\else $m_t$\fi}
\def\mc{\ifmmode m_c\else $m_c$\fi}
\def\mz{\ifmmode M_Z\else $M_Z$\fi}
\def\mw{\ifmmode M_W\else $M_W$\fi}
\def\mws{\ifmmode M_W^2 \else $M_W^2$\fi}
\def\mhs{\ifmmode M_H^2 \else $M_H^2$\fi}
\def\mzs{\ifmmode M_Z^2 \else $M_Z^2$\fi}
\def\mts{\ifmmode m_t^2 \else $m_t^2$\fi}
\def\mcs{\ifmmode m_c^2 \else $m_c^2$\fi}
\def\mchs{\ifmmode M_{H^\pm}^2 \else $M_{H^\pm}^2$\fi}
\def\ztwo{\ifmmode Z_2\else $Z_2$\fi}
\def\zone{\ifmmode Z_1\else $Z_1$\fi}
\def\mtwo{\ifmmode M_2\else $M_2$\fi}
\def\mone{\ifmmode M_1\else $M_1$\fi}
\def\tb{\ifmmode \tan\beta \else $\tan\beta$\fi}
\def\xw{\ifmmode x\sub w\else $x\sub w$\fi}
\def\ch{\ifmmode H^\pm \else $H^\pm$\fi}
\def\lum{\ifmmode {\cal L}\else ${\cal L}$\fi}
\def\inpb{\ifmmode {\rm pb}^{-1}\else ${\rm pb}^{-1}$\fi}
\def\infb{\ifmmode {\rm fb}^{-1}\else ${\rm fb}^{-1}$\fi}
\def\epem{\ifmmode e^+e^-\else $e^+e^-$\fi}
\def\ppb{\ifmmode \bar pp\else $\bar pp$\fi}

\newskip\zatskip \zatskip=0pt plus0pt minus0pt
\def\matth{\mathsurround=0pt}

\def\atversim#1#2{\lower0.7ex\vbox{\baselineskip\zatskip\lineskip\zatskip
  \lineskiplimit 0pt\ialign{$\matth#1\hfil##\hfil$\crcr#2\crcr\sim\crcr}}}

\def\Re{{\cal R \mskip-4mu \lower.1ex \hbox{\it e}\,}}
\def\Im{{\cal I \mskip-5mu \lower.1ex \hbox{\it m}\,}}

\def\sub#1{_{\lower.25ex\hbox{$\scriptstyle#1$}}}
\def\sul#1{_{\kern-.1em#1}}
\def\sll#1{_{\kern-.2em#1}}
\def\sbl#1{_{\kern-.1em\lower.25ex\hbox{$\scriptstyle#1$}}}
\def\ssb#1{_{\lower.25ex\hbox{$\scriptscriptstyle#1$}}}
\def\sbb#1{_{\lower.4ex\hbox{$\scriptstyle#1$}}}

\def\to{\rightarrow}

\def\rb{\ifmmode R_b\else $R_b$\fi}
\def\rc{\ifmmode R_c\else $R_c$\fi}
\def\ac{\ifmmode A_c\else $A_c$\fi}
\def\dmix{\ifmmode D^0-\bar D^0 \else $D^0$-$\bar D^0$\fi}
\def\dm{\ifmmode \Delta M_D \else $\Delta M_D$\fi}
\def\rb{\ifmmode R_b\else $R_b$\fi}
\def\mh{\ifmmode m\sbl H \else $m\sbl H$\fi}
\def\mch{\ifmmode M_{H^\pm} \else $M_{H^\pm}$\fi}
\def\mt{\ifmmode m_t\else $m_t$\fi}
\def\mc{\ifmmode m_c\else $m_c$\fi}
\def\mz{\ifmmode M_Z\else $M_Z$\fi}
\def\mw{\ifmmode M_W\else $M_W$\fi}
\def\mws{\ifmmode M_W^2 \else $M_W^2$\fi}

\def\mhs{\ifmmode m_H^2 \else $m_H^2$\fi}
\def\mzs{\ifmmode M_Z^2 \else $M_Z^2$\fi}
\def\mts{\ifmmode m_t^2 \else $m_t^2$\fi}
\def\mcs{\ifmmode m_c^2 \else $m_c^2$\fi}
\def\mchs{\ifmmode m_{H^\pm}^2 \else $m_{H^\pm}^2$\fi}
\def\ztwo{\ifmmode Z_2\else $Z_2$\fi}
\def\zone{\ifmmode Z_1\else $Z_1$\fi}
\def\mtwo{\ifmmode M_2\else $M_2$\fi}
\def\mone{\ifmmode M_1\else $M_1$\fi}
\def\bsg{\ifmmode b\to s\gamma\else $b\to s\gamma$\fi}
\def\tb{\ifmmode \tan\beta \else $\tan\beta$\fi}
\def\xw{\ifmmode x\sub w\else $x\sub w$\fi}
\def\ch{\ifmmode H^\pm \else $H^\pm$\fi}
\def\lum{\ifmmode {\cal L}\else ${\cal L}$\fi}
\def\inpb{\ifmmode {\rm pb}^{-1}\else ${\rm pb}^{-1}$\fi}
\def\infb{\ifmmode {\rm fb}^{-1}\else ${\rm fb}^{-1}$\fi}
\def\epem{\ifmmode e^+e^-\else $e^+e^-$\fi}
\def\ppb{\ifmmode \bar pp\else $\bar pp$\fi}

\def\be{\begin{equation}}
\def\ee{\end{equation}}

\newcommand{\bey}{\begin{eqnarray}}
\newcommand{\eey}{\end{eqnarray}}
\newcommand{\bdm}{\begin{displaymath}}
\newcommand{\edm}{\end{displaymath}}
\newcommand{\nn}{\nonumber}

\begin{document}
\vspace{3.0cm}
\preprint{\vbox {\vspace{2cm}
\hbox{WSU--HEP--0707}
}}

\vspace*{4cm}

\title{\boldmath Lifetime difference in $D^0$-${\overline D}^0$ mixing
within R-parity-violating SUSY}

\author{Alexey A.\ Petrov\vspace{5pt}}
\email{apetrov@wayne.edu}
\affiliation{Department of Physics and Astronomy\\[-6pt]
        Wayne State University, Detroit, MI 48201}
\affiliation{Michigan Center for Theoretical Physics\\[-6pt]
University of Michigan, Ann Arbor, MI 48109\\[-6pt] $\phantom{}$ }

\author{Gagik K. Yeghiyan\vspace{5pt}}
\email{ye_gagik@wayne.edu}
\affiliation{Department of Physics and Astronomy\\[-6pt]
        Wayne State University, Detroit, MI 48201}

\begin{abstract}
\vskip 0.2in We re-examine constraints from the recent evidence for
observation of the lifetime difference in $\DDbar$ mixing on the
parameters of supersymmetric models with $R$-parity violation (RPV).
We find that RPV SUSY can give large negative contribution to the
lifetime difference. We also discuss the importance of the choice of
weak or mass basis when placing the constraints on RPV-violating
couplings from flavor mixing experiments.
\end{abstract}

\def\thepage{{}}
\maketitle
\def\thepage{\arabic{page}}

\section{Introduction}
\renewcommand{\theequation}{1.\arabic{equation}}

Meson-antimeson mixing is an important vehicle for indirect studies of
New Physics (NP). Due to the absence of tree-level flavor-changing
neutral current transitions in the Standard Model (SM), it can only occur
via quantum effects associated with the SM and NP particles.
In fact, the
existence of both charm and top quark were inferred from the kaon and beauty
mixing amplitudes \cite{48}.
The estimates of masses of those particles were later
found to be in agreement with direct observations. This motivates indirect
searches for NP particles in a meson-antimeson mixing.

Recently, there has been a considerable interest in the only available
meson-antimeson mixing in the up-quark sector, the $\DDbar$ mixing
\cite{36}.
The fact that the search is indirect and complimentary to existing constraints
from the bottom-quark sector actually provides parameter space constraints
for a large variety of NP models~\cite{23,6}.

A flurry of recent experimental activity in that field led to the
observation of $\DDbar$ mixing from several different experiments
such as BaBar~\cite{34}, Belle~\cite{35} and CDF~\cite{CDFmix}. These results
have been combined by the Flavor Averaging Group (HFAG)~\cite{38}
to yield
\beq
y^{exp}_{\rm D} = (6.6 \pm 2.1) \cdot 10^{-3} \label{HFAG}
\label{1.1}
\eeq
\beq
x^{exp}_{\rm D} = 8.7^{+ 3.0}_{-3.4} \cdot 10^{-3},
\label{1.2}
\eeq
where $x_{\rm D}$ and $y_{\rm D}$ are defined as
\beq
x_{\rm D} \equiv {\Delta M_{\rm D} \over \Gamma_{\rm D}}, \qquad {\rm and}
\qquad y_{\rm D} \equiv {\Delta \Gamma_{\rm D} \over 2\Gamma_{\rm D}}\ \ ,
\label{xy}
\eeq
$\Gamma_{\rm D}$ is the average width of the two
neutral $D$ meson mass eigenstates, and $\Delta M_{\rm D}$,
$\Delta \Gamma_{\rm D}$ are the mass and width differences of the neutral
D-meson mass eigenstates. In the limit of CP-conservation,
$\Delta \Gamma_{D} \equiv \Gamma_+ - \Gamma_-$, where "+" and "-" are CP-even
and CP-odd D-meson eigenstates respectively.

One can also write $y_D$ as an absorptive part of the $\DDbar$ mixing
matrix~\cite{Petrov:2003un},
\begin{equation}\label{y1}
y_D=\frac{1}{\Gamma_{\rm D}}\sum_n \rho_n
\langle  \overline{D}^0| {\cal H}_w^{\Delta C=1}| n \rangle
\langle n | {\cal H}_w^{\Delta C=1}| D^0 \rangle,
\end{equation}
where $\rho_n$ is a phase space function that corresponds to a
charmless intermediate state $n$.  This relation shows that
$\Delta \Gamma_{\rm D}$ is driven by transitions $D^0, {\overline D}^0 \to n$,
{\it i.e.} physics of the $\Delta C=1$ sector.

Eqs.~(\ref{1.1}) and (\ref{1.2}) imply one-sigma window for the HFAG values
of $x_D$ and $y_{\rm D}$,
\bey
&&5.3 \cdot 10^{-3} < x_D < 11.7 \cdot 10^{-3}  \label{xbnds}
\qquad {\rm (one-sigma\ window)} \ \  \\
&& 4.5 \cdot 10^{-3} < y_{\rm D} < 8.7 \cdot 10^{-3}
\qquad \ \ {\rm (one-sigma\ window)} \ \
\label{ybnds}
\eey
In principle, these results can be used to constrain parameters of NP models
with the anticipated improved accuracy for the future D-mixing measurements.
In reality, those results can only provide the ballpark estimate to be used
for constraining NP models. The reason is that the SM estimate for the
parameters $x_{\rm D}$ and $y_{\rm D}$ is rather uncertain, as it
is dominated by long-distance QCD effects~\cite{29}-\cite{49}.
It was nevertheless shown that even this estimate provides rather stringent
constraints on the NP parameter space for many models affecting the
mass difference $x_{\rm D}$~\cite{23}, \cite{41}-\cite{46}.

It was recently shown~\cite{6} that $\DDbar$ mixing is a rather unique system, where
$y_{\rm D}$ can also be used to constrain the models of New Physics\footnote{A similar
effect is possible in the bottom-quark sector~\cite{Badin:2007bv}.}. This stems
from the fact that there is a well-defined theoretical limit (the flavor $SU(3)$-limit)
where the SM contribution vanishes and the lifetime difference is dominated
by the NP $\Delta C = 1$ contributions. In real world, flavor $SU(3)$ is, of
course, broken, so the SM contribution is proportional to a (second) power of
$m_s/\Lambda$, which is a rather small number. If the NP contribution to $y_{\rm D}$
is non-zero in the flavor $SU(3)$-limit, it can provide a large contribution
to the mixing amplitude.

To see this, consider a $D^0$ decay amplitude
which includes a small NP contribution, $A[D^0 \to n]=A_n^{\rm (SM)} + A_n^{\rm (NP)}$.
Experimental data for D-meson decays are known to be in a decent agreement
with the SM estimates \cite{47, 28}. Thus, $A_n^{\rm (NP)}$ should be smaller
than (in sum) the current theoretical and experimental
uncertainties in predictions for these decays.

One may rewrite equation~(\ref{y1}) in the form (neglecting the effects of
CP-violation)
\begin{eqnarray}
y_D &=& \sum_n \frac{\rho_n}{\Gamma_{\rm D}}
A_n^{\rm (SM)} \bar A_n^{\rm (SM)}
+ 2\sum_n \frac{\rho_n}{\Gamma_{\rm D}}
A_n^{\rm (NP)} \bar A_n^{\rm (SM)} +
\sum_n \frac{\rho_n}{\Gamma_{\rm D}}
A_n^{\rm (NP)} \bar A_n^{\rm (NP)} \ \ .
\label{approx}
\end{eqnarray}
The first term in this equation corresponds to the SM contribution, which vanishes in the
$SU(3)$ limit. In ref. \cite{6} the last term in (\ref{approx})
has been neglected, thus the NP
contribution to $y_{\rm D}$ comes there solely from the second term, due to
interference of $A_n^{\rm (SM)}$ and $A_n^{\rm (NP)}$.
While this contribution is in general non-zero in the flavor $SU(3)$
limit, in a large class of (popular) models it actually is~\cite{6, 37}. Then,
in this limit, $y_{\rm D}$ is completely dominated by pure $A_n^{\rm (NP)}$
contribution given by the last term in eq.~(\ref{approx})! It is clear that
the last term in equation (\ref{approx}) needs more detailed and careful studies,
at least within some of the NP models.

Indeed, in reality, flavor $SU(3)$ symmetry is broken, so the first term in
Eq.~(\ref{approx}) is not zero. It has been argued~\cite{29} that in fact
the SM $SU(3)$-violating contributions could be at a percent level, dominating the
experimental result. The SM predictions of $y_D$, stemming from evaluations of
long-distance hadronic contributions, are rather uncertain.
While this precludes us from placing explicit constraints on
parameters of NP models, it has been argued that, even in this situation,
an upper bound on the NP contributions can be placed~\cite{23} by displaying
the NP contribution only, i.e. as if there were no SM contribution at all.
This procedure is similar to what was traditionally done in the studies of
NP contributions to $K^0-\overline{K}^0$ mixing, so we shall employ it here too.

The purpose of this paper is to revisit the problem of the NP contribution to
$y_{\rm D}$ and provide constraints on R-parity-violating supersymmetric
(SUSY) models as a primary example. It has been recently argued in
\cite{17} that within \slash{\hspace{-0.25cm}R}- SUSY models, new physics
contribution to $y_{\rm D}$ is rather small, mainly because of stringent
constraints on the relevant pair products of RPV coupling constants.
However, this result has been
derived neglecting the transformation of these couplings from the weak
isospin basis to the quark mass basis. This approach seems to be quite
reasonable for the scenarios with the baryonic number violation.
However, in the scenarios with the leptonic number violation,
transformation of the RPV couplings from the weak eigenbasis to the quark
mass eigenbasis turns to be crucial, when applying the existing
phenomenological constraints on these couplings.

We show in the present paper that within R-parity-breaking
supersymmetric models with the leptonic number violation, new physics contribution to the
lifetime difference in $\DDbar$ mixing may be large, due to the last term
in eq.~(\ref{approx}). When being large, it is negative
(if neglecting CP-violation), i.e. opposite in sign
to what is implied by the recent experimental evidence for $\DDbar$
mixing.

The paper is organized as follows. In Section~2 we discuss the R-parity
violating interactions that, in particular, contribute to
$\DDbar$ lifetime difference. We
confront the form of these interactions in the weak isospin basis to
that in the quark mass basis, emphasizing the important differences.
In Section~3 we re-derive formulae for the RPV SUSY contribution to
$y_{\rm D}$. Unlike ref. \cite{17}, transformation of the RPV coupling
constants from the weak to the quark mass eigenbasis is taken into
account. Also the behavior of different \slash{\hspace{-0.25cm}R}- SUSY
contributions in the limit of the flavor SU(3) symmetry is discussed in
details. In Section~4 we examine the existing phenomenological constraints
on the RPV coupling constants. The importance of taking into account the
transformation of these couplings from the weak
to the mass eigenbasis is emphasized again. We present our numerical results in
Section~5. We conclude in Section~6. Appendices contain
some details of derivation of bounds on the pair products of RPV couplings,
relevant for our analysis.

\section{R-Parity Breaking Interactions: Weak vs Mass Eigenbases}

\setcounter{equation}{0}
\renewcommand{\theequation}{2.\arabic{equation}}

We consider a general low-energy supersymmetric scenario with no
assumptions made
on a SUSY breaking mechanism at the unification scales
$(\sim~(10^{16}~-~10^{18})GeV)$.
The most general Yukawa superpotential for an explicitly broken R-parity
supersymmetric theory is given by
\beq
W_{\slash{\hspace{-0.2cm}R}} = \sum_{i, j, k} \left[
\frac{1}{2} \lambda_{ijk} L_i L_j E^c_k +
\lambda^\prime_{ijk} L_i Q_j D^c_k +
\frac{1}{2} \lambda^{\prime \prime}_{ijk} U^c_i D^c_j D^c_k
\right] \label{2.1}
\eeq
where $L_i$, $Q_j$ are $SU(2)_L$ weak isodoublet lepton and quark
superfields, respectively; $E_i^c$, $U_i^c$, $D_i^c$ are SU(2) singlet
charged lepton, up- and down-quark superfields, respectively;
$\lambda_{ijk}$ and $\lambda^\prime_{ijk}$ are lepton number violating
Yukawa couplings, and $\lambda^{\prime \prime}_{ijk}$ is a baryon number
violating Yukawa coupling; $\lambda_{ijk} = - \lambda_{jik}$,
$\lambda^{\prime \prime}_{ijk} = - \lambda^{\prime \prime}_{ikj}$.
To avoid rapid proton decay, we assume that $\lambda^{\prime \prime}_{ijk} = 0$
and work with a lepton number violating \slash{\hspace{-0.25cm}R}- SUSY model.

For meson-to-antimeson oscillation processes, to the lowest order in the
perturbation theory, only the second term of (\ref{2.1}) is of the importance.
The relevant R-parity breaking part of the Lagrangian is the following:
\bey
\nn
{\cal L}_{\slash{\hspace{-0.2cm}}R} = \sum_{i ,j, k} \lambda^\prime_{ijk}
\Biggl[ - \widetilde{e}_{i_L} \bar{d}_{k_R}^w u_{j_L}^w -
\widetilde{u}_{j_L}^w \bar{d}_{k_R}^w e_{i_L} -
\widetilde{d}^{w^*}_{k_R} \bar{e}_{i_R}^c u_{j_L}^w +
\widetilde{\nu}_{i_L} \bar{d}_{k_R}^w d_{j_L}^w + \\
+ \widetilde{d}_{j_L}^w \bar{d}_{k_R}^w \nu_{i_L} +
\widetilde{d}^{w^*}_{k_R}
\bar{\nu}_{i_R}^c d_{j_L}^w \Biggr] + h.c. \label{2.2}
\eey
The superscript $w$ indicates that the quark and squark
states in (\ref{2.2}) are weak isospin eigenstates. The weak and
mass quark eigenstates are related by the unitary transformations
(we assume that left- and right-chiral quarks have the same transformation
matrices):
\beq
u^w_j = S_{u_{jn}} u_n, \hspace{0.5cm}
d^w_k = S_{d_{km}} d_m \label{2.3}
\eeq
where
\beq
\sum_{j, j^\prime} S^*_{u_{j^\prime n^\prime}} Y_{u_{j^\prime j}}
S_{u_{jn}} = \delta_{n n^\prime } h_{u_n}, \hspace{0.5cm}
\sum_{k, k^\prime} S^*_{d_{k^\prime m^\prime}} Y_{d_{k^\prime k}}
S_{d_{km}} = \delta_{m m^\prime } h_{d_m} \label{2.4}
\eeq
and
\beq
\sum_k S^*_{u_{kj}} S_{d_{kn}} = V_{jn} \label{2.5}
\eeq
In (\ref{2.4}) $Y_u$, $Y_d$ are quark-Higgs-quark R-parity conserving
Yukawa couplings
in the weak isospin basis and $h_u$, $h_d$ are these couplings in the
quark mass eigenbasis. In (\ref{2.5}), $V_{jn}$ stands as usually for the
(Standard Model) CKM matrix.

Generally speaking, squark transformation matrices from
the weak to the mass
eigenstates are different from those for quarks. Nevertheless,
we choose
for squarks to be rotated by the same matrices $S_u$ and $S_d$
that make quark mass matrices diagonal, i.e.
\bey
\nn
&& \widetilde{u}^w_{j_L} = S_{u_{jn}} \widetilde{u}_{n_L}, \hspace{0.5cm}
 \widetilde{u}^w_{j_R} = S_{u_{jn}} \widetilde{u}_{n_R}  \\
&& \widetilde{d}^w_{k_L} = S_{d_{km}} \widetilde{d}_{m_L}, \hspace{0.5cm}
 \widetilde{d}^w_{k_R} = S_{d_{km}} \widetilde{d}_{m_R} \label{2.6}
\eey
This is a super-CKM basis, in which the squark mass matrices
are non-diagonal and result in mass insertions that change the squark
flavors~\cite{30}-\cite{33}. This source of flavor violation is very
important in the pure MSSM sector. In particular, it plays crucial role
in examining the MSSM contribution to $D^0 - \bar{D}^0$ mass difference~\cite{23}.


In the R-parity breaking part of SUSY
Lagrangian,
flavor changing neutral currents are present {\it a priori}.
In order to simplify our analysis, we put all the squark masses to be nearly
equal.
Then the squark mass matrix is proportional to the identity matrix, i.e. it is
diagonal in any basis.

Using (\ref{2.3}), (\ref{2.5}) and (\ref{2.6}),
one may rewrite (\ref{2.2}) as
\bey
\nn
&& {\cal L}_{\slash{\hspace{-0.2cm}}R} = - \sum_{i, j, k, m, n, r}
\lambda^\prime_{ijk} S^*_{d_{km}} S_{d_{jn}} V_{rn}^*
\left[\widetilde{e}_{i_L} \bar{d}_{m_R} u_{r_L} +
\widetilde{u}_{r_L} \bar{d}_{m_R} e_{i_L} +
\widetilde{d}^{*}_{m_R} \bar{e}_{i_R}^c u_{r_L} \right] + \\
&& + \sum_{i, j, k, m, n} \lambda^\prime_{ijk} S_{d_{km}}^* S_{d_{jn}}
\left[\widetilde{\nu}_{i_L} \bar{d}_{m_R} d_{n_L} +
\widetilde{d}_{n_L} \bar{d}_{m_R} \nu_{i_L} +
\widetilde{d}^{*}_{m_R}
\bar{\nu}_{i_R}^c d_{n_L} \right] + h.c. \label{2.7}
\eey
At this point one may redefine, without loss of generality, the couplings
$\lambda^\prime$ as
\beq
\lambda^\prime_{ijk} S^*_{d_{km}} S_{d_{jn}} \to \lambda^\prime_{inm}
\label{2.8}
\eeq
This is also equivalent to choosing the weak and mass eigenbases
for down-quarks being the same, while for up-quarks they are
related by the CKM matrix\footnote{This definition of $\lambda^\prime$
is not unique. For example, Allanach et al.~\cite{1} used the up-quark
weak and mass eigenbases to be the same, relating the bases for down-quarks
by the CKM matrix. Another possibility is to redefine $\lambda^\prime$ in such a way
that (s)up-(s)down-charged (s)lepton vertices have the
couplings $\lambda^\prime$
while (s)down-down-(s)neutrino vertices have the couplings
$\lambda^\prime \cdot V_{CKM}$ \cite{4}.
Clearly all these approaches are equivalent.}.

Defining $ \widetilde{\lambda}^\prime_{irm} \ = \
V^*_{rn} \ \lambda^\prime_{inm}$ and renaming the summation indices,
we rewrite (\ref{2.7}) as
\bey
\nn
&& {\cal L}_{\slash{\hspace{-0.2cm}}R} = - \sum_{i, j, k}
\widetilde{\lambda}^\prime_{ijk}
\left[\widetilde{e}_{i_L} \bar{d}_{k_R} u_{j_L} +
\widetilde{u}_{j_L} \bar{d}_{k_R} e_{i_L} +
\widetilde{d}^{*}_{k_R} \bar{e}_{i_R}^c u_{j_L} \right] + \\
&& + \sum_{i, j, k} \lambda^\prime_{ijk}
\left[\widetilde{\nu}_{i_L} \bar{d}_{k_R} d_{j_L} +
\widetilde{d}_{j_L} \bar{d}_{k_R} \nu_{i_L} +
\widetilde{d}^{*}_{k_R}
\bar{\nu}_{i_R}^c d_{j_L} \right] + h.c. \label{2.10}
\eey
As it follows from (\ref{2.10}), (s)down-down-(s)neutrino vertices
have the weak eigenbasis couplings $\lambda^\prime$, while charged
(s)lepton-(s)down-(s)up vertices have the up quark mass eigenbasis
couplings $\tilde{\lambda}^\prime$.

Very often in the literature (see e.g. \cite{6},~\cite{17},~\cite{2}-\cite{5})
one neglects the difference between
$\lambda^\prime$ and $\widetilde{\lambda}^\prime$, based on the fact
that diagonal elements of the CKM matrix dominate over non-diagonal
ones, i.e.
\beq
V_{jn} = \delta_{jn} + O(\lambda) \qquad \mbox{so} \qquad
\widetilde{\lambda}_{ijk} \approx \lambda^\prime_{ijk} +
O(\lambda)
\label{2.11}
\eeq
where $\lambda = \sin\theta_c \sim 0.2$, with $\theta_c$ being the Cabibbo angle.

Notice that relation Eq.~(\ref{2.11}) is valid if only there is
{\it no hierarchy} in couplings $\lambda^\prime$. On the other hand, the existing
strong bounds on pair products $\lambda^\prime \times \lambda^\prime$ (or
$\widetilde{\lambda}^\prime \times \widetilde{\lambda^\prime}$)
\cite{1,2,3} and relatively loose bounds on individual couplings
$\lambda^\prime$~\cite{1} suggest that such a hierarchy may exist. As we
will see in Section~4, pair products
$\widetilde{\lambda}^\prime \times \widetilde{\lambda^\prime}$
may be orders of magnitude greater than corresponding products
$\lambda^\prime \times \lambda^\prime$.

To the end of this section, we explicitly
write down the terms of the R-parity breaking part of the
Lagrangian that contribute to $D^0 - \bar{D}^0$ lifetime difference:
\bey
\nn
{\cal L}{^{D^0 - \bar{D}^0}} = - \sum_i \Biggl[
\widetilde{\lambda}^\prime_{i21} \tilde{e}_{i_L} \bar{d}
\left(\frac{1 - \gamma_5}{2}\right)
c + \widetilde{\lambda}_{i22}^\prime \tilde{e}_{i_L}
\bar{s} \left(\frac{1 - \gamma_5}{2}\right) c + \\
\nn
+ \widetilde{\lambda}_{i11}^{\prime *}
\tilde{e}_{i_L}^{ *} \bar{u} \left(\frac{1 + \gamma_5}{2}\right)
d +  \widetilde{\lambda}_{i12}^{\prime *}
 \tilde{e}_{i_L}^{ *} \bar{u}
\left(\frac{1 + \gamma_5}{2} \right) s \Biggr] - \\
\nn
- \sum_k \Biggl[
\widetilde{\lambda}^\prime_{12k} \tilde{d}^{ *}_{k_R}
\bar{e}^c \left(\frac{1 - \gamma_5}{2} \right) c +
\widetilde{\lambda}^\prime_{22k} \tilde{d}^{ *}_{k_R} \
\bar{\mu}^c \left(\frac{1 - \gamma_5}{2} \right) c + \\
+ \widetilde{\lambda}^{\prime *}_{11k} \tilde{d}_{k_R}
\bar{u} \left(\frac{1 + \gamma_5}{2} \right) e^c +
\widetilde{\lambda}^{\prime *}_{21k} \tilde{d}_{k_R}
\bar{u} \left( \frac{1 + \gamma_5}{2} \right) \mu^c
\Biggr] \label{2.12}
\eey
In the next section we will integrate out heavy degrees
of freedom in (\ref{2.12}), thus finding
\slash{\hspace{-0.25cm}R}-SUSY part of $\Delta C = 1$
effective Hamiltonian. Then we will compute R-parity
breaking SUSY contribution to $\Delta \Gamma_D$.

\section{$D^0 - \bar{D}^0$ Lifetime Difference}
\setcounter{equation}{0}
\renewcommand{\theequation}{3.\arabic{equation}}

\begin{figure}[t]
\centerline{
\includegraphics[width=12cm]{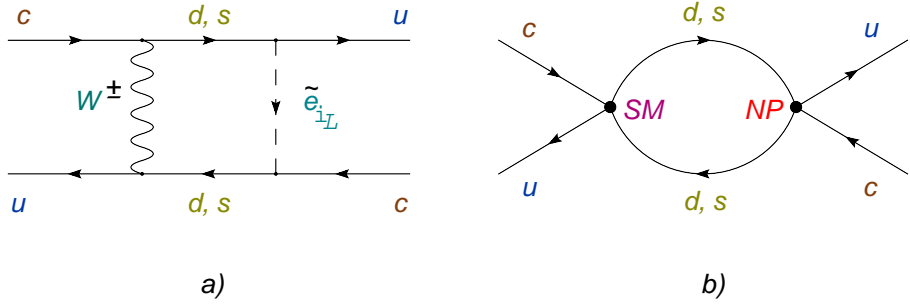}}
\vspace*{0.1cm}
\caption{\footnotesize $D^0 - \bar{D}^0$ mixing
diagrams with R-parity breaking interactions: a) within the
full electroweak theory; b) within the low-energy
effective theory. In these diagrams,
$D^0 - \bar{D}^0$ oscillations occur via two
subsequent $\Delta C =1$ transitions with the exchange of
$W^{\pm}$ boson and a charged "left" slepton, i=1,2,3.
}
\label{f1}
\end{figure}

\begin{figure}[t]
\centerline{
\includegraphics[width=12cm]{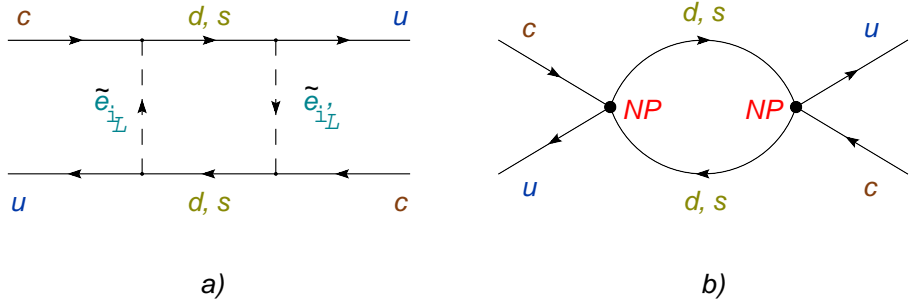}}
\vspace*{0.1cm}
\caption{\small
Same as in Fig.~\ref{f1}, however both of $\Delta C = 1$
transitions
are due to a charged slepton exchange now, $i=1,2,3$,
$i^\prime = 1,2,3$. Both of the effective $\Delta C = 1$
vertices are NP vertices.}
\label{f2}
\end{figure}

\begin{figure}[t]
\centerline{
\includegraphics[width=12cm]{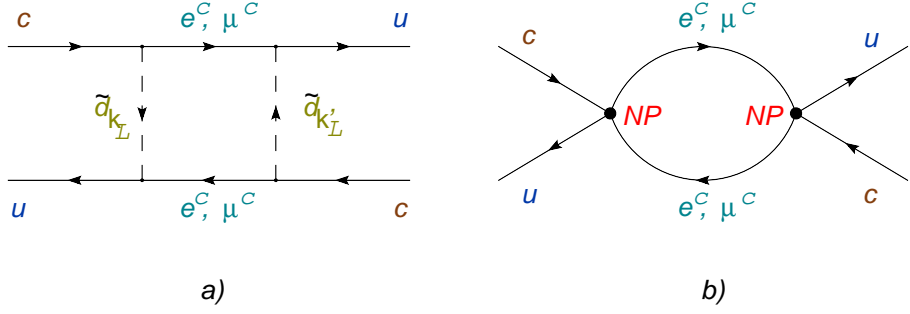}}
\vspace*{0.1cm}
\caption{\small
Same as in Fig.'s~\ref{f1},~\ref{f2}, however
both of $\Delta C = 1$ transitions occur due to exchange of
down-type
squarks now, $k=1,2,3$, $k^\prime = 1,2,3$.
Subsequently the intermediate charmless states
are charged (anti)lepton states.}
\label{f3}
\end{figure}

Assuming CP-conservation, the normalized $D^0 - \bar{D}^0$ lifetime
difference is given by
\beq
y_D = \frac{1}{2 m_D
\Gamma_D} Im \left[ \langle \bar{D}^0 | i \int d^4 x \
T\left\{ H_W^{\Delta C = 1} (x) H_W^{\Delta C = 1}(0)
\right\}|D^0 \rangle \right], \label{3.1}
\eeq
where $H_W^{\Delta C = 1}$ is an effective Hamiltonian including
both SM and NP parts. To the lowest order in the perturbation theory,
\slash{\hspace{-0.25cm}R}-SUSY contribution to
$D^0 - \bar{D}^0$ mixing comes from the one-loop graphs with
\begin{itemize}\addtolength{\itemsep}{-0.5\baselineskip}
\item{$W^\pm$ boson, charged slepton and two
down-type quarks (Fig.~\ref{f1}a);}
\item{two charged sleptons and two
down-type quarks (Fig.~\ref{f2}a);}
\item{two down-type squarks and
two charged leptons\footnote{As it follows from (\ref{2.12}),
lepton propagators in Fig.~\ref{f3}
must be constructed by contractions of
charge conjugates of the electron and/or muon field
operators.} (Fig.~\ref{f3}a) .}
\end{itemize}
Within the low-energy effective theory,
$D^0 - \bar{D}^0$ lifetime difference occurs as a result of
a bi-local transition with two $\Delta C = 1$ effective
vertices. The relevant low-energy diagrams in
Fig.'s~1b)~-~3b) are derived by integrating out of heavy
$W^\pm$ boson, charged slepton and down-type squark
degrees of freedom.

For R-parity-violating SUSY models one can therefore write
\beq
H_W^{\Delta C = 1} = H_{W_{SM}}^{\Delta C = 1}  +
H_{W_{\tilde{\ell}}}^{\Delta C = 1} +
H_{W_{\tilde{q}}}^{\Delta C = 1} \label{3.2}
\eeq
The first term in the r.h.s of (\ref{3.2}) is the
Standard Model contribution, whereas the second term comes
from $\Delta C = 1$ transitions with a slepton exchange
and the last term comes
from $\Delta C = 1$ transitions with a squark exchange.
The Standard model part of $\Delta C = 1$ effective
Hamiltonian is given by
\bey
\nn
H_{W_{SM}}^{\Delta C = 1} &=& \frac{G_F}{\sqrt{2}} \Biggl[
C_1 (\mu_c) \ \delta^{a_1 a_4} \ \delta^{a_3 a_2} +
C_2 (\mu_c) \ \delta^{a_1 a_2} \
\delta^{a_3 a_4} \Biggr]
\\ &\times&
\sum_{q_1,\ q_2} V_{u q_1} V^*_{c q_2} \ \bar{u}^{a_1}(x)
\gamma^\mu (1 - \gamma_5) q_1^{a_2}(x) \
\bar{q}_2^{a_3}(x) \gamma_\mu (1 - \gamma_5) c^{a_4}(x)
\label{3.3}
\eey
where $q_1 = s,d$, $q_2 = s,d$, $a_i$ are the color indices, and
$C_1$ and $C_2$ are the operator Wilson coefficients. The Wilson
coefficients are to be evaluated at a low-energy scale $\mu_c$, which
we choose here as $\mu_c = m_c$.

%
%

To simplify the following calculations, let us assume
that all the sleptons and all squarks are nearly degenerate, i.e.
\beq
m_{\tilde{e}_i} = m_{\tilde{\nu}_i} = m_{\tilde{\ell}},
\qquad \mbox{and} \qquad
m_{\tilde{d}_k} = m_{\tilde{u}_k} = m_{\tilde{q}}.
\label{3.4}
\eeq
With this assumption, the low energy effective Hamiltonian for the
R-parity-violating interactions are given by
\bey
\nn
H_{W_{\tilde{\ell}}}^{\Delta C = 1} &=& - \Biggl[
\widetilde{C}_1 (\mu_c) \ \delta^{a_1 a_4} \ \delta^{a_3 a_2} +
\widetilde{C}_2 (\mu_c) \ \delta^{a_1 a_2} \
\delta^{a_3 a_4} \Biggr]
\\ &\times&
\sum_{q_1, \ q_2} \frac{\lambda_{q_1 q_2}}{4 m_{\tilde{\ell}}^2}
\ \bar{u}^{a_1}(x)
(1 + \gamma_5) q_1^{a_2}(x) \
\bar{q}_2^{a_3}(x) (1 - \gamma_5) c^{a_4}(x),
\label{3.6}
\eey
and
\beq
H_{W_{\tilde{q}}}^{\Delta C = 1} = -
\sum_{\ell_1, \ \ell_2} \frac{\lambda_{\ell_1 \ell_2}}
{4 m_{\tilde{q}}^2}
\ \bar{u}^{a}(x)
(1 + \gamma_5) \ell_1^{c}(x) \
\bar{\ell}_2^{c}(x) (1 - \gamma_5) c^{a}(x)
\label{3.7}
\eeq
where $q_1 = s,d$, $q_2 = s,d$, $\ell_1 = e, \mu$, and
$\ell_2 = e, \mu$. The superscript $''c''$ stands for charge conjugation.
Also,
\beq
\lambda_{q_1 q_2} \ \equiv \ \sum_{i} \
\widetilde{\lambda}^{\prime *}_{i 1 q_1} \
\widetilde{\lambda}^{\prime}_{i 2 q_2}
\qquad\mbox{and}\qquad
\lambda_{\ell_1 \ell_2} \ \equiv \ \sum_{k} \
\widetilde{\lambda}^{\prime *}_{\ell_1 1 k} \
\widetilde{\lambda}^{\prime}_{\ell_2 2 k}
\label{3.8}
\eeq
We assume that $\lambda_{q_1 q_2}$ and $\lambda_{\ell_1, \ell_2}$ are real.

The insertions of Hamiltonians of eqs.~(\ref{3.3}), (\ref{3.6}), and (\ref{3.7}) can lead to
the lifetime difference in $\DDbar$ system. Let us write it as
\beq
y_D = y_{SM} + y_{SM,NP} + y_{\tilde{\ell} \tilde{\ell}} +
y_{\tilde{q} \tilde{q}},
\label{3.10}
\eeq
where
\bey
\nn
y_{SM, NP} = \frac{1}{2 m_D
\Gamma_D} \ Im \Biggl[ \langle \bar{D}^0 | i \int d^4 x \
T\Biggl\{H_{W_{SM}}^{\Delta C = 1} (x)
H_{W_{\tilde{\ell}}}^{\Delta C = 1}(0) +  \\
+ H_{W_{\tilde{\ell}}}^{\Delta C = 1} (x)
H_{W_{SM}}^{\Delta C = 1}(0)
\Biggr\}|D^0 \rangle \Biggr] \label{3.11}
\eey
is the term coming form the interference of the SM and NP contributions
to $H_W^{\Delta C = 1}$,  and
\beq
y_{\tilde{\ell} \tilde{\ell}} = \frac{1}{2 m_D
\Gamma_D} \ Im \Biggl[ \langle \bar{D}^0 | i \int d^4 x \
T\Biggl\{H_{W_{\tilde{\ell}}}^{\Delta C = 1} (x)
H_{W_{\tilde{\ell}}}^{\Delta C = 1}(0)
\Biggr\}|D^0 \rangle \Biggr],
\label{3.12}
\eeq
\beq
y_{\tilde{q} \tilde{q}} = \frac{1}{2 m_D
\Gamma_D} \ Im \Biggl[ \langle \bar{D}^0 | i \int d^4 x \
T\Biggl\{H_{W_{\tilde{q}}}^{\Delta C = 1} (x)
H_{W_{\tilde{q}}}^{\Delta C = 1}(0)
\Biggr\}|D^0 \rangle \Biggr] \label{3.13}
\eeq
are coming from two insertions of the NP vertices.

It might seem unreasonable to include double insertions of the
NP Hamiltonian to compute $y_D$, as each insertion generates a
contribution that is suppressed by some NP scale $M_{NP}$,
which in general is greater than the electroweak scale set here by $M_W$.
Yet, as the Standard Model contribution is zero in the
flavor SU(3) limit (i.e. suppressed by powers of strange quark
mass), New Physics contributions can be large~\cite{6}.
Also, as can be seen from refs.~\cite{6} and \cite{17}, $y_{SM,NP}$
resulting from the single insertion of the NP Hamiltonian is
forbidden in the $SU(3)$ flavor symmetry limit. Thus, double
insertion of the NP Hamiltonian can be important, especially if
this contribution does not vanish in the $SU(3)$ limit! This construction
can give numerically large contribution to $y_D$ if
$\left(M_W/M_{NP}\right)^2 > \left(m_s/m_c\right)^2$.

Note that contribution to $\Delta \Gamma_D$ is nonzero if
the intermediate states are the on-mass-shell
real physical states. It is therefore easy to see from the
energy-momentum conservation that diagrams like those
in Fig.'s~\ref{f1}-\ref{f3} but with b-quarks,
$\tau\tau$, $\tau\mu$ pairs running a loop, are
irrelevant for our analysis. While the diagrams with a
$\tau e$ pair running in a loop do give nonzero contribution
to $\Delta \Gamma_D$, their contributions are suppressed
by the available phase space. Thus, we shall not consider them too.

It is known that correlation function in (\ref{3.1}) (as well as those in
(\ref{3.11})-(\ref{3.13})) may be presented as a sum of
local $\Delta C = 2$ operators, which corresponds to $1/m_c$ power
expansion of (\ref{3.1}) (or (\ref{3.11}) - (\ref{3.13})). Here we
are interested in the lowest order terms in this expansion. Keeping
only the leading terms in $x_s \equiv m_{s}^2/m_c^2$ and
$x_d \equiv m_{d}^2/m_c^2$, we get
\bey
\nn
y_{SM,NP} = - \frac{G_F}{\sqrt{2}} \ \frac{\left(
K_1 + K_2 \right) }{4 \pi m_D \Gamma_D } \
\left(\frac{m_c^2}{m_{\tilde{\ell}}^2} \right)
 \Biggl[
\lambda_{sd} \sqrt{x_s x_d} \ + \\
+ \ \lambda
\left(\lambda_{ss} x_s - \lambda_{dd} x_d
\right) -
\lambda^2 \lambda_{ds} \sqrt{x_s x_d} \
\Biggr] \langle Q \rangle \label{3.14}
\eey
and
\bey
\nn
y_{\tilde{\ell} \tilde{\ell}} = \frac{m_c^2 \left(
\lambda_{ss}^2 + \lambda_{dd}^2 + 2
\lambda_{sd} \lambda_{ds} \right)}{192 \pi m_D
\Gamma_D m_{\tilde{\ell}}^4} \ \Biggl\{ - \left[
\frac{\widetilde{K}_2}{2} + \widetilde{K}_1
\right] \langle Q \rangle + \\
+ \ \left[
\widetilde{K}_2 - \widetilde{K}_1 \right]
\langle Q_S \rangle \Biggr\} \label{3.15}
\eey
where $\lambda = \sin{\theta_C}$ is the Wolfenstein parameter, and
\bey
\langle Q \rangle \equiv \langle \bar{D}^0 | \
\bar{u}^{a_1}(0) \gamma^\mu
\left(\frac{1 - \gamma_5}{2} \right)
c^{a_1}(0) \ \bar{u}^{a_2}(0) \gamma_\mu
\left(\frac{1 - \gamma_5}{2} \right)
c^{a_2}(0) \ | D^0 \rangle \\ \label{3.16}
\langle Q_S \rangle \equiv
\langle \bar{D}^0 | \ \bar{u}^{a_1}(0)
\left(\frac{1 + \gamma_5}{2}\right) c^{a_1}(0) \
\bar{u}^{a_2}(0) \left(\frac{1 + \gamma_5}{2} \right) \
c^{a_2}(0) \ | D^0 \rangle \label{3.17}
\eey
are the matrix elements of the effective low energy $\Delta C = 2$ operators and
\bey
&&
K_1 \ = \ 3 \ C_1 \ \widetilde{C}_1 + C_1 \ \widetilde{C}_2 +
C_2 \ \widetilde{C}_1,
\hspace{0.5cm} K_2 \ = \ C_2 \ \widetilde{C}_2 \label{3.18} \\
&& \widetilde{K}_1 \ = \ 3 \ \widetilde{C}_1^2 + 2 \
\widetilde{C}_1 \
\widetilde{C}_2, \hspace{2.3cm}
\widetilde{K}_2 \ = \ \widetilde{C}_2^2
\label{3.19}
\eey
are the Wilson coefficients. It is important to stress that $y_{SM, NP}$,
just like a Standard Model contribution, vanishes in the limit of exact
flavor $SU(3)$ symmetry - it is proportional to light quark masses via
$x_s$, $x_d$ and $\sqrt{x_s x_d}$. On the contrary, $y_{\tilde{\ell} \tilde{\ell}}$
is nonzero even in the limit of exact flavor $SU(3)$ symmetry! Therefore,
as we shall see in Section~5, $y_{\tilde{\ell} \tilde{\ell}}$ dominates over
$y_{SM, NP}$ if R-parity breaking coupling products
$\lambda_{ss}$ and/or $\lambda_{dd}$ approach their
boundaries. In other words, contribution of diagrams
in Fig.~2 with both of $\Delta C =1$ vertices generated
by new physics interactions, dominates over the contribution
of diagrams in Fig.~1, with one of the $\Delta C =1$
vertices coming from the Standard Model and the other
one coming from new physics.

Similarly, keeping only the leading order terms in
$x_e \equiv m_{e}^2/m_c^2$,
$x_\mu \equiv m_\mu^2/m_c^2$, one gets
\beq
y_{\tilde{q} \tilde{q}} =  \frac{- m_c^2 \left(
\lambda_{\mu \mu}^2 + \lambda_{ee}^2 + 2 \
\lambda_{\mu e} \lambda_{e \mu} \right)}
{192 \pi m_D
\Gamma_D \ m_{\tilde{q}}^4} \
\left[ \langle Q \rangle
\ + \  \langle Q_S \rangle \right].
\label{3.20}
\eeq
As one can see from (\ref{3.20}), $y_{\tilde{q} \tilde{q}}$ is
non-vanishing in the limit of exact flavor $SU(3)$ symmetry as well.

As usual, we parameterize matrix elements $\langle
Q \rangle$ and $\langle Q_s \rangle$ in terms of
B-factors~\cite{23}, i.e.
\beq
\langle Q \rangle = \frac{2}{3} \ f_D^2 \ m_D^2 \
B_D,
\hspace{0.5cm}
\langle Q_S \rangle = - \frac{5}{12} \ f_D^2 \ m_D^2
\ \bar{B}_D^S \label{3.21}
\eeq
where
\beq
\bar{B}_D^S \equiv B_D^S \ \frac{m_D^2}{m_c^2} \label{3.22}
\eeq
We shall follow the approach of ref.~\cite{6} and neglect
QCD running of the local $\Delta C = 1$ operators generated
by NP interactions. Thus, $\widetilde{C}_1 = 0$ and
$\widetilde{C}_2 = 1$, or
\beq
K_1 = C_1(m_c), \hspace{0.5cm} K_2 = C_2(m_c), \hspace{0.5cm}
\widetilde{K}_1 = 0, \hspace{0.5cm}
\widetilde{K}_2 = 1. \label{3.23}
\eeq
Using (\ref{3.21}) and (\ref{3.23}), one may rewrite
(\ref{3.14}), (\ref{3.15}) and (\ref{3.20}) in a following
form:
\bey
\nn
y_{SM,NP} = \frac{- \ G_F}{\sqrt{2}} \ \frac{f_D^2 B_D
m_D }{6 \pi \Gamma_D } \
\left(\frac{m_c^2}{m_{\tilde{\ell}}^2} \right) \
\left[C_1(m_c) + C_2(m_c) \right]
 \Big[
\lambda_{sd} \sqrt{x_s x_d} + \\
+ \ \lambda
\left(\lambda_{ss} x_s - \lambda_{dd} x_d
\right) -
\lambda^2 \lambda_{ds} \sqrt{x_s x_d} \ \Big] \label{3.24}
\eey
\beq
y_{\tilde{\ell} \tilde{\ell}} = \frac{- \ m_c^2 \
f_D^2 B_D m_D}{288 \pi
\Gamma_D \ m_{\tilde{\ell}}^4} \ \Biggl[\
\frac{1}{2} + \frac{5}{8} \frac{\bar{B}_D^S}{B_D} \
 \Biggr]
\left[\ \lambda_{ss}^2 + \lambda_{dd}^2 + 2 \
\lambda_{sd} \lambda_{ds} \ \right]
\label{3.25}
\eeq
\beq
y_{\tilde{q} \tilde{q}} =  \frac{m_c^2 \ f_D^2 B_D
m_D} {288 \pi \Gamma_D \ m_{\tilde{q}}^4} \
\Biggl[\ \frac{5}{8} \frac{\bar{B}_D^S}{B_D} -
1 \ \Biggr] \left[\ \lambda_{\mu \mu}^2 +
\lambda_{ee}^2 + 2 \
\lambda_{\mu e} \lambda_{e \mu} \ \right]
\label{3.26}
\eeq

Formulae (\ref{3.24})-(\ref{3.26}) involve only the lowest order
short-distance (perturbative) contribution to $\DDbar$ lifetime difference.
Yet, it has been mentioned already that long-distance effects play very
important role in $\DDbar$ oscillations. In particular, in the
Standard Model, where the short-distance contribution to $y_{\rm D}$
has a suppressing factor $\sim m_s^4/m_c^4$ \cite{24}, long distance
contribution to $\DDbar$ lifetime difference dominates \cite{29}.
However, within $\slash \hspace{-0.23cm} R$-SUSY models we have a
different
situation. As it is mentioned above, new physics contribution to
$y_{\rm D}$ is
non-vanishing in the exact flavor SU(3) limit, thus there is no
suppression in powers of $m_s/m_c$ in the dominant short-distance
NP terms. In what follows, long distance effects,
which may be interpreted as $\Lambda_{DCD}/m_c$ power corrections,
are subdominant. Thus, they may be neglected to the leading-order
approximation that is used throughout our paper.

Further analysis depends on bounds on R-parity breaking
coupling constants, so in the next section we discuss
the existing constraints on these couplings.

\section{Present Bounds on R-parity Breaking Coupling Constants}
\setcounter{equation}{0}
\renewcommand{\theequation}{4.\arabic{equation}}

Bounds on the R-parity violating couplings $\lambda^\prime$ have been widely
discussed in the literature \cite{1} - \cite{21}.
Summary of bounds on $\lambda^\prime_{ijk}$  may be found e.g. in~\cite{1}.
More recent (updated) bounds on some $\lambda^\prime \times \lambda^\prime$ pair
products, coming from the studies of $K^0 - \bar{K}^0$ and $B^0 - \bar{B}^0$ mixing
and $K^+ \to \pi^+ \nu \bar{\nu}$ decays, are presented in~\cite{2,5} and~\cite{18}
respectively.

It is interesting to note that bounds on RPV couplings coming from $K^0 - \bar{K}^0$
and $B^0 - \bar{B}^0$ mixing and empirical individual bounds on
couplings $\lambda^{\prime}_{ijk}$ are derived {\it neglecting} the difference
between $\lambda^\prime$ and $\tilde{\lambda}^\prime$. While for the individual bounds
it is a self-consistent approach, for the constraints on
RPV coupling pair products such an approach in general is questionable.

Empirical individual bounds on RPV couplings are derived, assuming that only one
coupling $\lambda^\prime_{ijk}$ is nonzero at a time. If such an assumption is made,
then it is easy to see that
\beq
\widetilde{\lambda}^\prime_{ijk} =
\lambda^\prime_{ijk} \times \left(1+
O(\lambda^2 = \sin^2{\theta_C}) \right), \\
\label{4.1}
\eeq
\beq
\widetilde{\lambda}^\prime_{ink} =
O(\lambda) \times \lambda^\prime_{ijk}
\label{4.2}
\eeq
if $n \neq j$, and
\beq
\widetilde{\lambda}^\prime_{rnm} = 0 \label{4.3}
\eeq
if $r \neq i$ or $m \neq k$.

Thus, as it follows from (\ref{4.1})-(\ref{4.3}),
when deriving an individual bound on
$\lambda^\prime_{ijk}$ by studying a given process,
there is no essential difference whether the
\slash{\hspace{-0.23cm}R}-SUSY diagram for this process
contains $\lambda^\prime_{ijk}$ or it contains
$\widetilde{\lambda}^\prime_{ijk}$ at the vertices.

Of course, in the realistic
\slash{\hspace{-0.23cm}R}-SUSY scenarios several
$\lambda^\prime$ couplings are in general non-zero. As it has
been pointed out in \cite{1}, even if at the unification
scales $(\sim~(10^{16} - 10^{18})$GeV)
one has only one non-zero RPV coupling, other non-zero
RPV couplings appear when evolving down from the unification
scales to the electroweak breaking scale. However, the individual
bounds on $\lambda^\prime$ couplings are still approximately valid, if
one assumes that one RPV coupling dominates over all other ones.
If several couplings dominate, individual bounds may still be used, if
they are not correlated or weakly correlated with each other.

The situation with the constraints on the RPV coupling pair products is
more complicated. As we will see,
bounds on $\tilde{\lambda}^\prime
\times \tilde{\lambda}^\prime$ and the corresponding
$\lambda^\prime \times \lambda^\prime$ products
may be different by several orders of
magnitude. One must therefore be careful when using the bounds
given in the literature and specify whether these bounds are
on $\lambda^\prime \times \lambda^\prime$ product or they are
on $\tilde{\lambda}^\prime \times \tilde{\lambda}^\prime$.
This may be easily done, using the following "rule of thumb":
\begin{itemize}\addtolength{\itemsep}{-0.5\baselineskip}
\item{If the process that is used to put constraints on the
RPV coupling products is described by diagram(s) with
down-down-sneutrino or down-sdown-neutrino vertices, bounds
are derived on a $\lambda^\prime \times \lambda^\prime$
product}.
\item{If such a process is described by diagram(s) with
up-down-charged slepton, up-sdown-charged lepton or
sup-down-charged lepton vertices, bounds
are derived on a
$\tilde{\lambda}^\prime \times \tilde{\lambda}^\prime$
product}.
\item{If both types of vertices are present, bounds are
derived on some admixture of
$\lambda^\prime \times \lambda^\prime$ and
$\tilde{\lambda}^\prime \times \tilde{\lambda}^\prime$
products.}
\end{itemize}

In addition to the individual bounds,
we use here constraints on the RPV coupling pair products that
are derived from study of $K^+ \to \pi^+ \nu \bar{\nu}$
decay and $\Delta m_{K^0}$. R-parity breaking SUSY contribution to
$K^+ \to \pi^+ \nu \bar{\nu}$
is described by tree-level diagrams with
a down-type squark exchange and quark-squark-neutrino interaction
vertices \cite{18,19,4}. Thus, this decay gives bounds on $\lambda^\prime
\times \lambda^\prime$ products.

The situation with $K^0 - \bar{K}^0$ mixing is more involved: there are
several sets of \slash{\hspace{-0.23cm}R}-SUSY diagrams that contribute
to this process. In order to get bounds on the RPV couplings, one assumes
that only a given RPV coupling product or a given sum of RPV coupling
products is nonzero. Possible bounds on the RPV coupling pair products
have been originally listed in \cite{3}. Recently these bounds have
been improved in \cite{2}.  Bounds that are relevant for our analysis
are presented in Appendix~A. We also specify which of them
are for $\lambda^\prime \times \lambda^\prime$
pair products and which of them are for
$\tilde{\lambda^\prime} \times \tilde{\lambda^\prime}$.

Keeping in mind everything that has been said above,
let us consider the RPV coupling products, which are
present in formulae (\ref{3.24})-(\ref{3.26}). We start with
\beq
\lambda_{ss} \equiv \sum_{i}{\
\widetilde{\lambda}^{\prime *}_{i12}
\widetilde{\lambda}^{\prime}_{i22}} = \sum_{i, j, n}{\ V_{1n}
V^*_{2j} \ \lambda^{\prime *}_{in2} \lambda^\prime_{ij2}}.
\label{4.4}
\eeq
Using Wolfenstein parametrization for the CKM matrix, keeping
for each $\lambda^\prime \times \lambda^{\prime *}$
product only the leading order term in $\lambda = \sin{\theta_C}$,
and assuming that all $\lambda^\prime \times \lambda^{\prime *}$
products are real (no new source of CP-violation), we
rewrite (\ref{4.4}) in a following form:
\bey
\nn
\lambda_{ss} \equiv \sum_{i}{\
\widetilde{\lambda}^{\prime *}_{i12} \
\widetilde{\lambda}^{\prime}_{i22}} = \sum_{i}{\
\lambda^{\prime *}_{i12} \ \lambda^\prime_{i22}} +
\lambda \Big[\sum_i{|\lambda^\prime_{i22}|^2} -
\sum_i{|\lambda^\prime_{i12}|^2} \Big] \\
\nonumber
+ \ A \lambda^2
\sum_{i}{\ \lambda^{\prime *}_{i12} \ \lambda^\prime_{i32}}
+ A \lambda^3 (1 + \rho - i \eta) \sum_{i}{\
\lambda^{\prime *}_{i32} \ \lambda^\prime_{i22}} \\
+ A^2 \lambda^5 (\rho - i \eta) \sum_i{
|\lambda^\prime_{i32}|^2}
\label{4.5}
\eey
There is a strong bound on the Cabibbo-favored term in
the r.h.s. of (\ref{4.5}) from the $K^+ \to \pi^+ \nu \bar{\nu}$ decay.
Assuming that $\lambda^{\prime *}_{i1k} \ \lambda^\prime_{i2k} \neq 0$
only for k=2, one gets \cite{18}
\beq
|\lambda^{\prime *}_{i12} \ \lambda^\prime_{i22}| \leq
6.3  \cdot 10^{-5} \left(\frac{m_{\tilde{q}}}{300GeV}
\right)^2 \label{4.6}
\eeq
We have rescaled the bound of ref. \cite{18} to the units of
$m_{\tilde{q}}/300$~GeV. Values of the squark masses less
than 300 GeV are disfavored by many experiments (see
\cite{15} for more details). For this reason, we follow
ref.~\cite{2} assuming that $m_{\tilde{q}} \geq 300$~GeV.

If squarks happen to be superheavy\footnote{We thank X.~Tata for
discussion of this scenario.}, there is still a strong bound on the
Cabibbo favored term in (\ref{4.5}) from $K^0 - \bar{K}^0$ mixing.
As it follows from our discussion in Appendix~A,
\beq
|\sum_{i}{\lambda^{\prime *}_{i12} \ \lambda^\prime_{i22}}|
\leq 2.7 \times 10^{-3} \left(\frac{m_{\tilde{\ell}}}{100GeV}
\right)^2 \label{4.7}
\eeq
Thus, the Cabibbo favored term in (\ref{4.5}) is strongly suppressed,
if one assumes that only $\lambda^\prime_{i12} \neq 0$.
and $\lambda^\prime_{i22} \neq 0$.
On the other hand, even  under such an assumption,
one still has
\bdm
\lambda_{ss} \equiv \widetilde{\lambda}^{\prime *}_{i12} \
\widetilde{\lambda}^{\prime}_{i22} \neq
\lambda^{\prime *}_{i12} \
\lambda^{\prime}_{i22}
\edm
due to the first order Cabibbo suppressed terms in (\ref{4.5}).
Furthermore, constraints (\ref{4.6}) or (\ref{4.7}) may in
particular be satisfied, when $|\lambda^\prime_{i22}|$
is close to its boundary value whereas
$|\lambda^\prime_{i12}|  \to 0$, and vice versa. Taking
into account that individual bounds are, in general, orders
of magnitude looser than (\ref{4.6}) or (\ref{4.7}), it is
not hard to see that $\lambda_{ss}$ is dominated by the
first order Cabibbo suppressed term in (\ref{4.5}).

Further on we will very often deal with a situation, when
expanding $\widetilde{\lambda}^\prime \times
\widetilde{\lambda}^\prime$ products in a basis of
$\lambda^\prime$ couplings, the Cabibbo favored term is
negligible whereas the first order Cabibbo suppressed term
dominates, and the only possible constraints on the
first order Cabibbo suppressed term are the individual
bounds on $\lambda^\prime$ couplings. In order to use these
bounds we assume hereafter that only one coupling
$\lambda^\prime_{ijk}$ dominates at a time.

After making such an assumption, it is easy to see that
\bey
\nn
- 0.025 \left(\frac{m_{\tilde{q}}}{300GeV} \right)^2 \leq
\lambda_{ss}
\leq 0.29, \quad \mbox{if} \quad m_{\tilde{q}} \leq 1 TeV, \\
- 0.29 \leq \lambda_{ss}
\leq 0.29, \quad \mbox{if} \quad m_{\tilde{q}} \geq 1 TeV \label{4.8}
\eey
The upper bound on $\lambda_{ss}$ is derived when one of
$\lambda^\prime_{i22}$ couplings dominates. Individual bounds on
$\lambda^\prime_{i22}$ are the loosest for $i=3$~\cite{1}.
For $m_{\tilde{q}} \geq 300$GeV,
$|\lambda_{322}|~\leq~1.12$ - this
is the perturbativity bound on $\lambda_{322}$. The lower
bound on $\lambda_{ss}$ is derived when one of
$\lambda^{\prime}_{i12}$ couplings dominates. Individual bounds on
$\lambda^\prime_{i12}$ are the loosest for i=3 again:
$|\lambda^\prime_{312}| \leq 0.33 (m_{\tilde{q}}/300GeV)$,
if $m_{\tilde{q}} \leq 1 TeV$ and
$|\lambda_{312}| \leq 1.12$ - the perturbativity bound,
if $m_{\tilde{q}} \geq 1 TeV$.

It is important to stress that, in general, as it follows
from (\ref{4.6}), (\ref{4.7}), (\ref{4.8}),
\beq
\lambda_{ss} \equiv \sum_{i}{
\widetilde{\lambda}^{\prime *}_{i12}
\widetilde{\lambda}^{\prime}_{i22}} \gg
\sum_{i}{\lambda^{\prime *}_{i12}
\lambda^{\prime}_{i22}} \label{4.9}
\eeq
Thus, as it has been already pointed out in the
beginning of this section, bounds on
$\tilde{\lambda}^\prime \times \tilde{\lambda}^\prime$
products differ by several orders of magnitude from those on
corresponding $\lambda^\prime \times \lambda^\prime$ products.
In the considered case,
$\tilde{\lambda}^\prime \times \tilde{\lambda}^\prime$
product is restricted by much weaker bound than corresponding
$\lambda^\prime \times \lambda^\prime$ product.

Relation (\ref{4.9}) plays crucial
role in our analysis. We will see in the next section that,
as a consequence of this relation, R-parity breaking
SUSY contribution to $\Delta \Gamma_D$ is quite large.

For $\lambda_{dd}$, analysis is performed in exactly the same
way and yields
\bey
\nn
- 0.025 \left(\frac{m_{\tilde{q}}}{300GeV} \right)^2 \leq
\lambda_{dd}
\leq 0.29, \quad if \quad m_{\tilde{q}} \leq 1 TeV, \\
- 0.29 \leq \lambda_{dd}
\leq 0.29, \quad if \quad m_{\tilde{q}} \geq 1 TeV \label{4.10}
\eey
Also, the relation similar to (\ref{4.9}) is obtained:
\beq
\lambda_{dd} \equiv \sum_{i}{
\widetilde{\lambda}^{\prime *}_{i11}
\widetilde{\lambda}^{\prime}_{i21}} \gg
\sum_{i}{\lambda^{\prime *}_{i11}
\lambda^{\prime}_{i21}} \label{4.11}
\eeq
and relation (\ref{4.11}) is as crucial as (\ref{4.9}).
It is also useful to transform (\ref{4.8}) and (\ref{4.10})
onto restrictions on $\lambda_{ss}^2$ and $\lambda_{dd}^2$:
\beq
\lambda_{ss}^2 \approx \lambda^2 \Big[
\sum_{i}{|\lambda^\prime_{i22}|^2} -
\sum_{i}{|\lambda^\prime_{i12}|^2} \Big]^2 \leq
0.0841 \label{4.22}
\eeq
\beq
\lambda_{dd}^2 \approx \lambda^2 \Big[
\sum_{i}{|\lambda^\prime_{i21}|^2} -
\sum_{i}{|\lambda^\prime_{i11}|^2} \Big]^2 \leq
0.0841 \label{4.23}
\eeq
Bounds on $\lambda_{ds}$ and $\lambda_{sd}$ are derived using
the experimental data for $\Delta m_{K^0}$. As it follows from
formula (\ref{A.1}) in Appendix~A,
\beq
|\lambda_{ds}| \equiv
\Big|\sum_{i} \widetilde{\lambda}^{\prime*}_{i11}
\widetilde{\lambda}^{\prime}_{i22}\Big|
\leq 1.7 \cdot 10^{-6}
\left(\frac{m_{\tilde{\ell}}}{100GeV}\right)^2 \label{4.12}
\eeq
In order to derive constraints
on $\lambda_{sd}$, one must write it in a following
form (using
$\lambda^\prime_{ijk} = V_{nj} \widetilde{\lambda}^\prime_{ink}$):
\beq
\lambda_{sd} \equiv \sum_{i}{
\widetilde{\lambda}^{\prime *}_{i12}
\widetilde{\lambda}^{\prime}_{i21}} =
\left( V^*_{11} V_{22} \right)^{-1} \Biggl[ \sum_{i}{\lambda^{\prime *}_{i12}
\lambda^{\prime}_{i21}} -
{\sum_{j,n}}^\prime{\ V^*_{j1} V_{n2} \Big(
\sum_i{\widetilde{\lambda}^{\prime *}_{ij2}
\widetilde{\lambda}^{\prime}_{in1}}} \Big) \Biggr]
\label{4.13}
\eeq
where prime indicates that the sum over $j$ and $n$ does not contain
the term with $j=1$ and $n=2$. Bounds on the terms present in r.h.s.
of (\ref{4.13}) are given in Appendix~A. Using these bounds, one
can see that
\beq
\lambda_{sd} < \mbox{few} \times 10^{-7}
\left(\frac{m_{\tilde{\ell}}}{100GeV}\right)^2
\label{4.14}
\eeq
It is interesting to note that such strong constraints on $\lambda_{ds}$
and on $\lambda_{sd}$ are derived assuming that only one
$\tilde{\lambda}^\prime \times \tilde{\lambda}^\prime$ or
$\lambda^\prime \times \lambda^\prime$ product is nonzero.
It is also assumed that pure MSSM sector
gives negligible contribution to $\Delta m_{K^0}$ \cite{2}.
These two assumptions are not necessarily true. If one gives
up these assumption, then destructive interference of the pure
MSSM and \slash{\hspace{-0.2cm}R}-SUSY diagrams or the one
of different \slash{\hspace{-0.2cm}R}-SUSY diagrams will
somehow distort bounds (\ref{4.13}), (\ref{4.14}). However,
unless there is a fine-tuning or an exact cancelation between
two (or more) diagram contributions, it is very unlikely for
the distortion of these bounds to be such that $\lambda_{ds}$
and/or $\lambda_{sd}$ be $\sim 10^{-1}$ or $\sim 10^{-2}$.
Therefore in our numerical calculations we will
use the following relations:
\bey
\lambda_{ds} \ll \lambda_{ss}, \lambda_{dd}
\label{4.15} \\
\lambda_{sd} \ll \lambda_{ss}, \lambda_{dd}
\label{4.16}
\eey
For the remaining four coupling products - $\lambda_{ee}$,
$\lambda_{\mu \mu}$, $\lambda_{\mu e}$ and
$\lambda_{e \mu}$ - that are contained in the expression (3.26)
for $y_{\tilde{q} \tilde{q}}$,
the analysis is
similar to that for $\lambda_{ss}$ and $\lambda_{dd}$.
For the details and subtleties
of the analysis, we refer the reader to Appendix~B. Here we only
point out that bounds on $\lambda_{ee}$,
$\lambda_{\mu \mu}$ are the following:
\bea
\nn
- 0.91 \cdot 10^{-3}
\left(\frac{m_{\tilde{q}}}{300GeV}\right)^2 \leq
&\lambda_{ee}& \leq 3.83 \cdot 10^{-3}
\left(\frac{m_{\tilde{q}}}{300GeV}\right)^2
\label{4.17} \\
- 0.0072
\left(\frac{m_{\tilde{q}}}{300GeV}\right)^2 \leq
&\lambda_{\mu \mu}& \leq 0.091
\left(\frac{m_{\tilde{q}}}{300GeV}\right)^2, \quad \mbox{if} \quad
m_{\tilde{q}} \leq 530 \mbox{~GeV}, \\
- 0.0072
\left(\frac{m_{\tilde{q}}}{300GeV}\right)^2 \leq
&\lambda_{\mu \mu}& \leq 0.29, \qquad\qquad\qquad\quad \mbox{if} \quad
m_{\tilde{q}} \geq 530 \mbox{~GeV}.
\label{4.18}
\eea
Also, for two other couplings we get
\bey
\nn
&& |\lambda_{\mu e}| \leq 0.019 \left(\frac{m_{\tilde{q}}}
{300GeV} \right)^2, \hspace{0.4cm}
|\lambda_{e \mu}| \leq 0.019 \left(\frac{m_{\tilde{q}}}
{300GeV} \right)^2,
\quad \mbox{if} \quad m_{\tilde{q}} \leq
530 \mbox{~GeV} \\
&& |\lambda_{\mu e}| \leq 0.033 \left(\frac{m_{\tilde{q}}}
{300GeV} \right), \hspace{0.5cm}
|\lambda_{e \mu}| \leq 0.033 \left(\frac{m_{\tilde{q}}}
{300GeV} \right),
\quad \mbox{if} \quad m_{\tilde{q}} \geq
530 \mbox{~GeV}
\label{4.19}
\eey
We also obtain that
\beq
\lambda_{\mu e} \approx \lambda_{e \mu} \label{4.20}
\eeq
As $m_{\tilde{q}}$ increases, squark mass dependent
empirical bounds on the RPV
couplings are replaced by squark mass independent
perturbativity bounds. In formulae
(\ref{4.17})-(\ref{4.19}), we indicate the change in the
behavior of the bounds with the squark mass, if it occurs
for $m_{\tilde{q}} \leq 1$TeV.

When transforming (\ref{4.17})-(\ref{4.20}) onto the
restrictions on $\lambda_{ee}^2$, $\lambda_{\mu \mu}^2$,
$\lambda_{\mu e} \lambda_{e \mu}$, one can see that these
restrictions are much weaker than the relevant
constraints listed in ref. \cite{17}. This is because in
the present paper we do not neglect the transformations
of RPV couplings from the weak eigenbasis to the
quark mass eigenbasis. More precisely, we do not neglect
the difference between $\widetilde{\lambda}^\prime
\times \widetilde{\lambda}^\prime$ and $\lambda^\prime
\times \lambda^\prime$ pair products.

From (\ref{4.17})-(\ref{4.20}), one can also see that
generally speaking,
\beq
\lambda_{\mu \mu}^2 \gg \lambda_{\mu e} \lambda_{e \mu}
\gg \lambda_{ee}^2 \label{4.21}
\eeq
It is worth mentioning here that additional bounds on
$\lambda_{ee}$, $\lambda_{\mu \mu}$, $\lambda_{\mu e}$, $\lambda_{e
\mu}$ may be derived from studying rare D-meson decays, such as $D
\to X \ell^+ \ell^-$, $D^0 \to \ell^+ \ell^-$, etc \cite{28}. As it
follows from the analysis performed in ref. \cite{28}, bounds
derived in this way may be even stronger than those given by
(\ref{4.17}) -(\ref{4.19}). Bounds coming from the rare D-meson
decays are however still to be elaborated in details, taking into
account new experimental data, as well as possible impact of the
long-distance SM and (short-distance) pure MSSM contributions. Such
an elaboration is beyond the scope of this paper, in particular
because $y_{\tilde{q} \tilde{q}}$ turns to be a (numerically)
subdominant part of the new physics contribution to $D^0 -
\bar{D}^0$ lifetime difference, even if we use constraints on
$\lambda_{ee}$, $\lambda_{\mu \mu}$, $\lambda_{\mu e}$, $\lambda_{e
\mu}$ given by (\ref{4.17})-(\ref{4.19}) (see the next section).

Having obtained constraints on all RPV coupling products in
(\ref{3.24})-(\ref{3.26}), we may proceed to computation
of $y_{SM, NP}$, $y_{\tilde{\ell} \tilde{\ell}}$,
$y_{\tilde{q} \tilde{q}}$.

\section{Numerical Analysis}
\setcounter{equation}{0}
\renewcommand{\theequation}{5.\arabic{equation}}

In our numerical calculations we use~\cite{15}
$G_F = 1.166 \cdot 10^{-5} ~\mbox{GeV}^{-2}$, $\lambda \approx 0.23$,
$\Gamma_D \approx 1.6 \cdot 10^{-12}$~GeV,
$m_D \approx 1.865$~GeV; $m_c \equiv m_c(m_c) \approx 1.25$~GeV,
$m_s(2GeV)\approx 95$~MeV,
\bdm
m_s(m_c) \approx m_s(2GeV) \left(\frac{\alpha_s(m_c)}
{\alpha_s(2GeV)}\right)^{12/25} \approx 105 \mbox{~MeV},
\quad x_s \equiv \frac{m_s^2(m_c)}{m_c^2(m_c)} \approx 0.007;
\edm
$C_1(m_c) = -0.411$, $C_2(m_c) \approx 1.208$ \cite{24},
$B_D \approx 0.8$ \cite{24,25}, $f_D \approx 0.22$~\cite{26}.

While the value of $B_D$ is known from the lattice QCD
calculations, there is no theoretical or experimental
prediction on $B_D^{S}$. Here we follow the approach of
ref. \cite{24}, assuming that
\beq
B_D^{S} = B_D, \quad B_D^{S} = 0.8 B_D, \quad
B_D^{S} = 1.2 B_D. \label{5.1}
\eeq
Let us first determine the sign of $y_{SM, NP}$,
$y_{\tilde{\ell} \tilde{\ell}}$, $y_{\tilde{q} \tilde{q}}$.
Using relations (\ref{4.15}), (\ref{4.16}), (\ref{4.21}),
one may rewrite equations (\ref{3.24})-(\ref{3.26}) in a much simpler
form,
\bey
y_{SM,NP} \ &\approx & \
\frac{- \ G_F}{\sqrt{2}} \ \frac{f_D^2 B_D
m_D }{6 \pi \Gamma_D } \
\left(\frac{m_c^2}{m_{\tilde{\ell}}^2} \right) \
\Big[ C_1(m_c) + C_2(m_c) \Big] \
\lambda \ \lambda_{ss} \ x_s
\label{5.2}
\\
y_{\tilde{\ell} \tilde{\ell}} \ &\approx & \
\frac{- \ m_c^2 \
f_D^2 B_D m_D}{288 \pi
\Gamma_D \ m_{\tilde{\ell}}^4} \ \Biggl[\
\frac{1}{2} + \frac{5}{8} \frac{\bar{B}_D^S}{B_D} \
 \Biggr]
\left[\ \lambda_{ss}^2 + \lambda_{dd}^2  \right]
\label{5.3}
\\
y_{\tilde{q} \tilde{q}} \ &\approx& \ \frac{m_c^2 \ f_D^2 B_D
m_D} {288 \pi \Gamma_D \ m_{\tilde{q}}^4} \
\Biggl[\ \frac{5}{8} \frac{\bar{B}_D^S}{B_D} -
1 \ \Biggr] \ \lambda_{\mu \mu}^2 \label{5.4}
\eey
It follows from (\ref{5.2}), (\ref{5.3}) that the sign of
$y_{SM, NP}$ is opposite to that of $\lambda_{ss}$ and
$y_{\tilde{\ell} \tilde{\ell}} < 0$.

One can see from (\ref{5.4}) that the sign of
$y_{\tilde{q} \tilde{q}}$ is determined by the factor
$\Big[\ \frac{5}{8} \frac{\bar{B}_D^S}{B_D} -
1 \ \Big]$. As it follows from
(\ref{3.22}) and (\ref{5.1}),
for $m_c \equiv m_c(m_c) \approx 1.25$GeV,
this factor is positive, hence
\bdm
y_{\tilde{q} \tilde{q}} > 0.
\edm
On the other hand,
$\Big[\ \frac{5}{8} \frac{\bar{B}_D^S}{B_D} -
1 \ \Big]$ and hence $y_{\tilde{q} \tilde{q}}$ flips its
sign when using the charm quark pole
mass~\footnote{To derive the proper value of $m_c^{pole}$, the
two-loop relation between the pole and
$\overline{MS}$ quark masses must be used. This is because the
$\overline{MS}$ value of the c-quark mass has been
extracted using the perturbative QCD analysis up to the
order $\alpha_s^2$ \cite{15}. One can check that
the use of the three loop relation between the pole and
$\overline{MS}$ quark masses \cite{27} leads to the physically
meaningless result $m_c^{pole} \approx 1.93 \mbox{~GeV} > m_D$.},
$m_c^{pole} \approx 1.65$~GeV.

In general, such an ambiguity in sign of $y_{\tilde{q} \tilde{q}}$
may cause a trouble in numerical evaluation of the results, signaling
the need for next-to-leading order evaluation of the appropriate
contributions, where the scheme ambiguity cancels out. Here we disregard
this sign ambiguity, as $y_{\tilde{q} \tilde{q}}$ turns to be a (numerically)
subdominant part of the new physics contribution to
$D^0 - \bar{D}^0$ lifetime difference. In our opinion, the use of
the $\overline{MS}$ charm mass, $m_c(m_c) = 1.25$~GeV, is more
appropriate in this calculation. Then $y_{\tilde{q} \tilde{q}}$ has
positive sign.

Let us proceed to our results. It is convenient
to start with $y_{\tilde{q} \tilde{q}}$. Using the listed
numerical values of parameters present in (\ref{5.4}), we
get
\bey
\nn
&& B_D^S = 0.8 B_D: \quad y_{\tilde{q} \tilde{q}}
\approx 0.0011 \ \lambda_{\mu \mu}^2 \left(\frac{300GeV}
{m_{\tilde{q}}}\right)^4 \\
&& B_D^S = B_D: \quad \quad \
y_{\tilde{q} \tilde{q}}
\approx 0.0038 \ \lambda_{\mu \mu}^2 \left(\frac{300GeV}
{m_{\tilde{q}}}\right)^4 \label{5.5}\\
\nn
&& B_D^S = 1.2 B_D: \quad y_{\tilde{q} \tilde{q}}
\approx 0.0064 \ \lambda_{\mu \mu}^2 \left(\frac{300GeV}
{m_{\tilde{q}}}\right)^4
\eey
As it follows from (\ref{5.5}), to the lowest order in
the perturbation theory,
$y_{\tilde{q} \tilde{q}}$ is highly
sensitive to the choice of
parameters $B_D^S$ and $B_D$. Moreover, if one uses the
approach of ref. \cite{17}, choosing $\bar{B}_D^S = B_D$
or $B_D^S = (m_c^2/m_D^2) B_D \approx 0.45 B_D$,
$y_{\tilde{q} \tilde{q}}$ flips the
sign\footnote{$y_{\tilde{q} \tilde{q}}$ is equivalent
to $- y_{(RPV-RPV,l)}$ in the notations of \cite{17}.}.

Using the bounds on $\lambda_{\mu \mu}$ given by
(\ref{4.18}) yields
\bey
\nn
&& B_D^S = 0.8 B_D: \quad y_{\tilde{q} \tilde{q}}
\leq 0.9 \cdot 10^{-5}  \\
&& B_D^S = B_D: \quad \quad \
y_{\tilde{q} \tilde{q}}
\leq 3.12 \cdot 10^{-5}  \label{5.6}  \\
\nn
&& B_D^S = 1.2 B_D: \quad y_{\tilde{q} \tilde{q}}
\leq 5.34 \cdot 10^{-5}
\eey
for $m_{\tilde{q}} \leq 530$~GeV and
\bey
\nn
&& B_D^S = 0.8 B_D: \quad y_{\tilde{q} \tilde{q}}
\leq 0.9 \cdot 10^{-5} \left(\frac{530GeV}
{m_{\tilde{q}}}\right)^4 \\
&& B_D^S = B_D: \quad \quad \
y_{\tilde{q} \tilde{q}}
\leq 3.12 \cdot 10^{-5} \left(\frac{530GeV}
{m_{\tilde{q}}}\right)^4 \label{5.7}\\
\nn
&& B_D^S = 1.2 B_D: \quad y_{\tilde{q} \tilde{q}}
\leq  5.34 \cdot 10^{-5} \left(\frac{530GeV}
{m_{\tilde{q}}}\right)^4
\eey
for $m_{\tilde{q}} \geq 530$~GeV.

Thus, if using bounds on $\lambda_{e e}$,
$\lambda_{\mu \mu}$, $\lambda_{\mu e}$,
$\lambda_{e \mu}$, given by (\ref{4.17}) - (\ref{4.20}),
one obtains that $y_{\tilde{q} \tilde{q}}$ is at least
by two orders of magnitude less than the experimental value of
$y_{\rm D}$. As it was mentioned above, constraints
on $\lambda_{e e}$,
$\lambda_{\mu \mu}$, $\lambda_{\mu e}$,
$\lambda_{e \mu}$ and hence
on $y_{\tilde{q} \tilde{q}}$ may become even stronger
if one elaborates the constraints on RPV couplings coming
from the rare $D$-meson decays. Further on we simply
disrespect $y_{\tilde{q} \tilde{q}}$ because of its
smallness. This way we also avoid the problems related
to the dependence of the obtained results on the choice of
the renormalization scheme and  $B_D$-factors.

Consider $y_{SM, NP}$ now. For this quantity one gets
\beq
y_{SM, NP} \approx 0.0040 \
\lambda_{ss} \left(\frac{100GeV}{
m_{\tilde{\ell}}} \right)^2 \label{5.8}
\eeq
which after using (\ref{4.8}) yields
\beq
- 0.0011 \left(\frac{100GeV}{
m_{\tilde{\ell}}} \right)^2 \leq
y_{SM, NP} \leq 0.99 \cdot 10^{-4} \left(\frac{m_{\tilde{q}}}{
300GeV} \right)^2 \left(\frac{100GeV}{
m_{\tilde{\ell}}} \right)^2 \label{5.9}
\eeq
for $m_{\tilde{q}} \leq 1$~TeV and
\beq
- 0.0011 \left(\frac{100GeV}{
m_{\tilde{\ell}}} \right)^2 \leq
y_{SM, NP} \leq 0.0011 \left(\frac{100GeV}{
m_{\tilde{\ell}}} \right)^2 \label{5.10}
\eeq
for $m_{\tilde{q}} \geq 1$~TeV.

As it follows from (\ref{5.9}), (\ref{5.10}),
$|y_{SM, NP}|$ may be by an order of magnitude greater
than it was quoted in \cite{17}\footnote{
$y_{SM, NP}=-y_{(SM-RPV)}$ in the notations of \cite{17}.}.
This is because the analysis in ref. \cite{17} has been restricted by
consideration of $m_{\tilde{q}} = 100$~GeV only. On the other hand, as
it follows from Table~I of ref. \cite{1} and our analysis
in Section~4, bounds on RPV couplings and hence on
$\lambda_{ss}$ become weaker for the greater values of
squark masses. Else, unlike ref.'s \cite{6,17},
we obtain that $y_{SM, NP}$ can be
both positive and negative. This is because, as one can
see from equation (\ref{4.5}) and the
following it discussion, $\lambda_{ss}$ may have both of
signs even if one assumes that all RPV couplings are
real and positive.

Finally, consider $y_{\tilde{\ell} \tilde{\ell}}$. Using the numerical
values of the parameters present in (\ref{5.3}), one gets
\bey
\nn
&& B_D^S = 0.8 B_D: \quad y_{\tilde{\ell} \tilde{\ell}} \approx
- 1.25 \left[\lambda_{ss}^2 + \lambda_{dd}^2 \right]
\left(\frac{100GeV}{m_{\tilde{\ell}}}\right)^4 \\
&& B_D^S = B_D: \quad \quad \
y_{\tilde{\ell} \tilde{\ell}} \approx
- 1.47 \left[\lambda_{ss}^2 + \lambda_{dd}^2 \right]
\left(\frac{100GeV}{m_{\tilde{\ell}}}\right)^4
\label{5.11} \\
\nn
&& B_D^S = 1.2 B_D: \quad y_{\tilde{\ell} \tilde{\ell}} \approx
- 1.69 \left[\lambda_{ss}^2 + \lambda_{dd}^2 \right]
\left(\frac{100GeV}{m_{\tilde{\ell}}}\right)^4
\eey
As one can see from (\ref{5.11}), varying the
ratio $B_D^S/B_D$ from 0.8 to 1.2, one gets
about 15\% uncertainty in the predictions for
$y_{\tilde{\ell} \tilde{\ell}}$. Thus,
$y_{\tilde{\ell} \tilde{\ell}}$ is only weakly sensitive to the
choice of the parameter $B_D^S$. As we are
interested in the order of the effect only, we may for a
simplicity assume $B_D^S = B_D$ hereafter.

To be consistent with a one dominant coupling approximation,
we will assume that only one of the coupling products
$\lambda_{ss}$ or $\lambda_{dd}$ is at its boundary at a
time. Notice however that if we allow both $\lambda_{ss}$
and $\lambda_{dd}$ to be simultaneously large, our results
will change at most by a factor two, which is inessential, if
one is interested in the order-of-magnitude of the effect only.

Using the bounds on $\lambda_{ss}^2$ and $\lambda_{dd}^2$
given by (\ref{4.22}) and (\ref{4.23}) we obtain
\beq
-0.12 \left(\frac{100GeV}{m_{\tilde{\ell}}}\right)^4
\leq y_{\tilde{\ell} \tilde{\ell}} < 0 \label{5.12}
\eeq
It is important to stress that $|y_{\tilde{\ell} \tilde{\ell}}|$
may be $\sim 10^{-1}$, if $m_{\tilde{\ell}} = 100$~GeV.

This result is in contradiction with the one of
ref.~\cite{17}: $y_{RPV-PRV,q} = - y_{\tilde{\ell} \tilde{\ell}}
\leq 2.5 \cdot 10^{-11}$, for $m_{\tilde{\ell}} = 100$GeV.
This contradiction is related to the
fact that authors of ref.~\cite{17}, following other papers
on the meson-antimeson mixing phenomenon, have neglected the
transformation of the RPV couplings from the weak eigenbasis
to the quark mass eigenbasis. This allowed them to impose very
stringent constraints on $\lambda_{ss}^2$ and $\lambda_{dd}^2$
from $K^+ \to \pi^+ \nu \bar{\nu}$ decay.
As it follows from our discussion in Section~4, this approach
is not always appropriate\footnote{Unless one imposes the conditions
$\lambda^\prime_{i22} \sim \lambda^\prime_{i12}$ and
$\lambda^\prime_{i21} \sim \lambda^\prime_{i11}$.}.

We are now able to compute the total New Physics contribution
to $D^0 - \bar{D^0}$ lifetime difference,
\bdm
y_{new} = y_{SM, NP} + y_{\tilde{\ell} \tilde{\ell}} +
y_{\tilde{q} \tilde{q}}.
\edm
As it is mentioned above, we neglect $y_{\tilde{q} \tilde{q}}$
because of its smallness. Also, as it follows from (\ref{5.8}) and
(\ref{5.11}), $y_{\tilde{\ell} \tilde{\ell}} \gg y_{SM, NP}$ unless
$\lambda_{dd} \to 0$ and the ratio
$\lambda_{ss}/m_{\tilde{\ell}}^2$ is small enough. It is not very
hard to see after doing some algebra that
\beq
-0.12 \left(\frac{100GeV}{m_{\tilde{\ell}}}\right)^4
\leq
y_{\tilde{\ell} \tilde{\ell}} + y_{SM, NP} \leq
2.72 \cdot 10^{-6} \label{5.13}
\eeq
The (negative) lower bound in (\ref{5.13}) is derived
neglecting $y_{SM, NP}$ as compared to
$y_{\tilde{\ell} \tilde{\ell}}$. The (positive) upper
bound in (\ref{5.13}) is derived for
$\lambda_{dd} = 0$ and $\lambda_{ss} = - 0.00136
\left( m_{\tilde{\ell}}/100GeV \right)^2$, when
$y_{SM, NP} = - 2 y_{\tilde{\ell} \tilde{\ell}}$.
As it follows from (\ref{5.6}) and (\ref{5.13}),
$y_{new}$ is negligible, if positive, and may be
as large as $\sim 10^{-1}$, if negative.

Thus, within the R-parity breaking supersymmetric models
with the lepton number violation, new physics contribution
to $D^0 - \bar{D}^0$ lifetime difference is
{\it predominantly negative} and may exceed in absolute
value the experimentally allowed interval. In order to
avoid a contradiction with the experiment, one must either
have a large positive contribution from the Standard Model, or
place severe restrictions on the values of RPV couplings.
As it follows from \cite{29}, $y_{SM}$ may be as large as
$\sim 1\%$. In what follows, $|y_{new}|$ must be
$\sim 1\%$ or smaller as well. If $|y_{new}| \sim 1\%$,
one may neglect $y_{SM, NP}$ as compared to
$y_{\tilde{\ell} \tilde{\ell}}$. Then, imposing condition
\beq
- 0.01 \leq y_{new} \approx
y_{\tilde{\ell} \tilde{\ell}} \label{5.14}
\eeq
one obtains that either $m_{\tilde{\ell}} > 185$GeV, or
if $m_{\tilde{\ell}} \leq 185$GeV, condition
(\ref{5.14}) implies new bounds on $\lambda_{ss}$ and
$\lambda_{dd}$:
\bey
|\lambda_{ss}| \leq 0.082 \left(\frac{m_{\tilde{\ell}}}{100GeV}
\right)^2 \label{5.15} \\
|\lambda_{dd}| \leq 0.082 \left(\frac{m_{\tilde{\ell}}}{100GeV}
\right)^2 \label{5.16}
\eey
Note that bounds (\ref{5.15}) and (\ref{5.16}) may not be
saturated simultaneously. (\ref{5.15}) is saturated if
$\lambda_{dd} = 0$. Subsequently, (\ref{5.16}) is saturated
if $\lambda_{ss} = 0$. For the opposite limiting case,
$\lambda_{ss} = \lambda_{dd}$, one gets $\sqrt{2}$ times
stronger restrictions:
\beq
|\lambda_{ss}| \leq 0.058 \left(\frac{m_{\tilde{\ell}}}{100GeV}
\right)^2, \hspace{1cm}
|\lambda_{dd}| \leq 0.058 \left(\frac{m_{\tilde{\ell}}}{100GeV}
\right)^2 \label{5.17}
\eeq
It is interesting to compare the restrictions on $\lambda_{ss}$
and $\lambda_{dd}$, given by (\ref{5.15})-(\ref{5.17}), with
those derived in \cite{23} from study of $D^0 - \bar{D}^0$
mass difference. Translated to our notations, we
may rewrite the relevant constraints of ref. \cite{23} in the
following form:
\beq
\lambda_{ss} \leq 0.085 \sqrt{x_{exp}} \left(\frac{m_{\tilde{q}}}
{500GeV} \right), \hspace{1cm}
\lambda_{dd} \leq 0.085 \sqrt{x_{exp}} \left(\frac{m_{\tilde{q}}}
{500GeV} \right) \label{5.18}
\eeq
This constraint has been derived assuming that
$m_{\tilde{q}} = m_{\tilde{\ell}}$. If
$m_{\tilde{q}} \neq m_{\tilde{\ell}}$, bounds in (\ref{5.18})
must be divided by the factor $\frac{1}{2} \sqrt{1 +
m_{\tilde{q}}^2/m_{\tilde{\ell}}^2}$, as it follows from
formulae (130)-(134) of ref. \cite{23}. Assuming for a
simplicity that $m_{\tilde{q}}^2 \gg m_{\tilde{\ell}}^2$ and
inserting $x_{exp} = 0.0117$ into (\ref{5.18}), one gets
\beq
\lambda_{ss} \leq 0.0037 \left(\frac{m_{\tilde{\ell}}}
{100GeV} \right), \hspace{1cm}
\lambda_{dd} \leq 0.0037 \left(\frac{m_{\tilde{\ell}}}
{100GeV} \right) \label{5.19}
\eeq

Thus, bounds of \cite{23} on $\lambda_{ss}$ and
$\lambda_{dd}$ are about 20 times stronger than our ones.
On the other hand, constraints of ref. \cite{23}
on the
RPV coupling products are derived in the limit when the pure
MSSM contribution to $\Delta m_D$ is negligible. Generally
speaking, the MSSM contribution to $D^0 - \bar{D}^0$ mass
difference is significant even for the squark masses being
about 2GeV. In what follows, the destructive interference of
the pure MSSM and \slash{\hspace{-0.23cm}R}-SUSY contributions
may distort bounds (\ref{5.19}), making them inessential as
compared to (\ref{5.15})-(\ref{5.17}) or even to
(\ref{4.8}), (\ref{4.10}).

Contrary to this, pure MSSM contributes to $\Delta \Gamma_D$
only in the next-to-leading order via two-loop dipenguin
diagrams. Naturally, this contribution is expected to be small.
In what follows, unlike those of ref. \cite{23},
our constraints on the RPV coupling products
$\lambda_{ss}$ and $\lambda_{dd}$, given by
(\ref{5.15})-(\ref{5.17}), seem to be
insensitive or weakly sensitive to assumptions on the
pure MSSM sector of the theory.

Thus, our main result is that within the R-parity breaking
supersymmetric theories with the leptonic number violation,
new physics contribution to $\Delta \Gamma_D$ may be quite
large and is predominantly negative.


For simplicity we assumed that all sleptons
have nearly the same mass and all squarks have nearly the same
mass. It is easy to see that
taking into account the difference between the slepton masses
does not affect our main results. There are however subtleties
concerning to the squark masses. First,
recall that our analysis has been performed for
$m_{\tilde{q}} \geq 300$~GeV. While this constraint is
quite reasonable for $\tilde{d}$ and $\tilde{s}$, bottom
squark is still allowed experimentally to be about
100~GeV~\cite{15}. On the other hand, we have seen that
bounds on $y_{SM, NP}$ and $y_{\tilde{\ell} \tilde{\ell}}$ either
grow or are insensitive to the squark masses. As for the
bound on $y_{\tilde{q} \tilde{q}}$, it is insensitive on
$m_{\tilde{q}}$ for low values of the squark masses. Thus,
no new effect is going to be observed, if one takes the
squark masses to be about 100GeV.

Another point to be made, is that the squark mass matrix
is in general non-diagonal in the super-CKM basis, if
one takes the squark masses to be different. In this
case, to take properly into account the squark mass insertion effects,
one should also give up the simplifying assumption
that left- and right-chiral quarks (of a same flavor) have
a same transformation matrix from the weak eigenbasis to the mass
eigenbasis. It has been
already mentioned in Section~2, that no new flavor violation
effects are obtained, however this may somehow weaken bounds
(\ref{4.17}) - (\ref{4.19})
on $\lambda_{ee}$, $\lambda_{\mu \mu}$,  $\lambda_{\mu e}$
$\lambda_{e \mu}$, when applying arguments analogous to
those used in Section~4. However, as it was mentioned above,
$\lambda_{ee}$, $\lambda_{\mu \mu}$,  $\lambda_{\mu e}$
$\lambda_{e \mu}$ are expected to get additional
strong constraints from the analysis of the rare $D$-meson decays,
so that one may expect for $y_{\tilde{q} \tilde{q}}$ to be
in any case restricted by even more stringent bound than (\ref{5.5}).
In other words, giving up the assumption of nearly equal squark masses
leads to complication of the analysis without observation
of any new effect. If being large, RPV SUSY contribution to the
lifetime difference in $\DDbar$ mixing still may have only
negative sign.

When studying the lifetime difference in  $\DDbar$ mixing within the Standard Model
and beyond, one usually assumes
that CP-violating effects are negligible \cite{6,29,24,37,17}. Following this
strategy, we have chosen for the RPV coupling products that contribute to
$\DDbar$ mixing amplitude to be real. The natural question arises if our
results may be affected by possible complex
phases of these coupling products.
Clearly, $|y_{new}|$ still may be large, however
the complex phases may possibly affect its sign.
One may suggest - because of
no evidence of CP-violation in
$\DDbar$ system \cite{34,35} -  that the phases of the
relevant RPV coupling products are small. In this case,
contribution to $\DDbar$ lifetime difference, proportional to the imaginary
parts of the RPV coupling products, is subdominant and cannot affect the
sign of $y_{new}$: if being large in the absolute value, $y_{new}$
is negative .
Yet, it may happen that RPV coupling products that contribute to $\DDbar$
mixing have large phases, and no evidence of CP-violation in $\DDbar$ system is
related to the fact that - unlike the $\DDbar$ oscillations -
\slash{\hspace{-0.23cm}R}-SUSY contribution to $D^0$
meson decays is rather small. In that case the formalism, used in our paper, is
not valid anymore. More general and involved approach should be used, taking
into account possible correlations in the values of $\DDbar$ mass and
lifetime differences as well as possible correlations in the SM, pure MSSM and
RPV sector contributions. Thus, to clarify if the RPV couplings complex phases
may affects the sign of the NP contribution to $\DDbar$ lifetime difference,
thorough and detailed study of the case,
when the relevant phases are large, is needed.

\section{Conclusion}

We computed a possible contribution from R-parity-violating
SUSY models to the lifetime difference in $\DDbar$ mixing.
Even though the $\DDbar$ system is rather unique in that
the Standard Model predicts vanishing of $y_D$ in a symmetry limit,
the technique and results described here can be applied to other
heavy flavored systems, especially those where the the Standard Model
predictions are very small, such as $B_d$-system. The contribution
from RPV SUSY models with the leptonic number violation
is found to be negative, i.e. opposite in sign
to what is implied by recent experimental evidence, and
possibly quite large, which implies stronger constraints on the
size of relevant RPV couplings.

We discussed currently available constraints on those couplings (especially
on the products of them), available from kaon mixing and rare kaon decays.
We emphasize that the use of these data in charm mixing has to be done
carefully separating the constraints on RPV couplings taken in the mass
and weak eigenbases, given the gauge and CKM structure of $\DDbar$ mixing
amplitudes.

\acknowledgments

Authors are grateful to S. Pakvasa and X. Tata for valuable discussions.

This work has been supported by the grants
NSF~PHY-0547794 and DOE~DE-FGO2-96ER41005.

\appendix

\section{Bounds on the RPV coupling pair products from
$\Delta m_{K^0}$}
\setcounter{equation}{0}
\renewcommand{\theequation}{A.\arabic{equation}}

R-parity breaking part of SUSY contributes to $K^0 - \bar{K}^0$ mixing
by the tree-level diagram with a sneutrino exchange, by the so-called
L2 type of box diagrams with $W^\pm$ boson and a charged slepton exchange
and by the so-called L4 type of box diagrams with all four vertices being
new physics generated vertices \cite{2}. Bounds on the RPV coupling
products are derived assuming that only a given pair product or a given
sum of pair products is non-zero.

Here we list the bounds, derived in \cite{2}, that are relevant for our
analysis. We consider only the case when the pair products are real.
We specify which of constraints are for
$\lambda^\prime \times \lambda^\prime$ products and which of them are
for $\tilde{\lambda}^\prime \times \tilde{\lambda}^\prime$:
\bey
&& |\lambda_{ds}| \equiv
\Big|\sum_{i} \widetilde{\lambda}^{\prime*}_{i11}
\widetilde{\lambda}^{\prime}_{i22}\Big|
\leq 1.7 \cdot 10^{-6}
\left(\frac{m_{\tilde{\ell}}}{100GeV}\right)^2
 \label{A.1} \\
&& \Big|\sum_{i} \widetilde{\lambda}^{\prime*}_{i32}
\widetilde{\lambda}^{\prime}_{i11}
\Big| \leq 2.2 \cdot 10^{-6}
\left(\frac{m_{\tilde{\ell}}}{100GeV}\right)^2
 \label{A.2} \\
&& \Big|\sum_{i} \widetilde{\lambda}^{\prime*}_{i32}
\widetilde{\lambda}^{\prime}_{i21} \Big|
\leq 5.1 \cdot 10^{-7}
\left(\frac{m_{\tilde{\ell}}}{100GeV}\right)^2
 \label{A.3} \\
&&\Big|\sum_{i} \widetilde{\lambda}^{\prime*}_{i12}
\widetilde{\lambda}^{\prime}_{i31} \Big|
\leq 7.5 \cdot 10^{-6}
\left(\frac{m_{\tilde{\ell}}}{100GeV}\right)^2
 \label{A.4} \\
&& \Big|\sum_{i} \widetilde{\lambda}^{\prime*}_{i22}
\widetilde{\lambda}^{\prime}_{i31} \Big|
\leq 3.3 \cdot 10^{-5}
\left(\frac{m_{\tilde{\ell}}}{100GeV}\right)^2
\label{A.5} \\
&&\Big|\sum_i{\lambda^{\prime *}_{i12}
\lambda^\prime_{i21}} \Big| \leq 9.8 \cdot 10^{-8}
\left(\frac{m_{\tilde{\ell}}}{100GeV}\right)^2
\label{A.6} \\
&&\Big|\sum_{i,k}{\lambda^{\prime *}_{i1k}
\lambda^\prime_{i2k}} \Big| \leq 2.7 \cdot 10^{-3}
\ for \ m_{\tilde{\ell}} = 100GeV, \
m_{\tilde{q}} = 300GeV \label{A.7}
\eey
If one assumes that the RPV coupling products are
non-zero only for a given $i$ and a given $k$, one
may apply them to each term in the above sums.

Bounds (\ref{A.1}) - (\ref{A.5}) are derived from charged slepton
mediated L2 diagrams and (\ref{A.6}) is derived from a tree level
sneutrino mediated diagram. Naturally these bounds scale with the
slepton mass squared. Contrary to this, to derive (\ref{A.7}), both
sneutrino mediated and squark mediated L4 diagrams are used. Thus,
it is not easy to scale this bound. However for $m_{\tilde{\ell}} =
100GeV$ and $m_{\tilde{q}} = 300GeV$, the squark mediated diagrams
contribution is about 10\% of that of the slepton mediated ones
\cite{2}. In what follows, (\ref{A.7}) is also approximately valid
if $m_{\tilde{q}} \gg m_{\tilde{\ell}}$. Then this bound may be
scaled with the slepton mass squared as well. Assuming that
$\lambda^{\prime *}_{i1k} \lambda^\prime_{i2k} \neq 0$ only for a
given value of k, one gets
\beq
\Big|\sum_{i}{\lambda^{\prime *}_{i1k}
\lambda^\prime_{i2k}} \Big| \leq 2.7 \cdot 10^{-3}
\left(\frac{m_{\tilde{\ell}}}{100GeV}\right)^2
\label{A.8}
\eeq
We do not use bounds of \cite{2} for $ij2 \times ij1$
combination products. Using our "rule of thumb" one can
see that these are bounds on some admixture of
$\lambda^{\prime *}_{ij2} \lambda^\prime_{ij1}$ and
$\widetilde{\lambda}^{\prime *}_{ij2}
\widetilde{\lambda}^\prime_{ij1}$.  We use instead
earlier bounds of ref. \cite{3}. These bounds are derived
using L2 diagrams only, neglecting L4 ones.
These diagrams vertices  contain
$\tilde{\lambda}^\prime$ couplings, but not
$\lambda^\prime$. Thus one has
\bey
\Big|\sum_i{\widetilde{\lambda}^{\prime *}_{i12}
\widetilde{\lambda}^{\prime}_{i11}}\Big| \leq
1.4 \cdot 10^{-6} \left(\frac{m_{\tilde{\ell}}}
{100GeV} \right)^2  \\ \label{A.9}
\Big|\sum_i{\widetilde{\lambda}^{\prime *}_{i22}
\widetilde{\lambda}^{\prime}_{i21}}\Big| \leq
1.4 \cdot 10^{-6} \left(\frac{m_{\tilde{\ell}}}
{100GeV} \right)^2  \\ \label{A.10}
\Big|\sum_i{\widetilde{\lambda}^{\prime *}_{i32}
\widetilde{\lambda}^{\prime}_{i31}}\Big| \leq
7.7 \cdot 10^{-4} \left(\frac{m_{\tilde{\ell}}}
{100GeV} \right)^2 \label{A.11}
\eey

\section{Bounds on $\lambda_{ee}$,
$\lambda_{\mu \mu}$, $\lambda_{e \mu}$,
$\lambda_{\mu e}$}
\setcounter{equation}{0}
\renewcommand{\theequation}{B.\arabic{equation}}

We may present $\lambda_{ee}$,
$\lambda_{\mu \mu}$, $\lambda_{\mu e}$,
$\lambda_{e \mu}$ in a following form:
\bey
\lambda_{ee} \equiv \sum_{k}{
\widetilde{\lambda}^{\prime *}_{11k}
\widetilde{\lambda}^{\prime}_{12k}} =
\sum_{k}{\lambda^{\prime *}_{11k} \lambda^{\prime}_{12k}} +
\lambda \left[\sum_{k} |\lambda^\prime_{12k}|^2 -
\sum_{k} |\lambda^\prime_{11k}|^2 \right] + O(\lambda^2)
\label{B.1} \\
\lambda_{\mu \mu} \equiv \sum_{k}{
\widetilde{\lambda}^{\prime *}_{21k}
\widetilde{\lambda}^{\prime}_{22k}} =
\sum_{k}{\lambda^{\prime *}_{21k} \lambda^{\prime}_{22k}} +
\lambda \left[\sum_{k} |\lambda^\prime_{22k}|^2 -
\sum_{k} |\lambda^\prime_{21k}|^2 \right] + O(\lambda^2)
\label{B.2} \\
\lambda_{\mu e} \equiv \sum_{k}{
\widetilde{\lambda}^{\prime *}_{11k}
\widetilde{\lambda}^{\prime}_{22k}} =
\sum_{k}{\lambda^{\prime *}_{11k} \lambda^{\prime}_{22k}} +
\lambda \left[\sum_{k} \lambda^{\prime *}_{12k}
\lambda^{\prime}_{22k} - \sum_{k} \lambda^{\prime *}_{11k}
\lambda^{\prime}_{21k}
 \right] + O(\lambda^2)
\label{B.3} \\
\lambda_{e \mu} \equiv \sum_{k}{
\widetilde{\lambda}^{\prime *}_{21k}
\widetilde{\lambda}^{\prime}_{12k}} =
\sum_{k}{\lambda^{\prime *}_{21k} \lambda^{\prime}_{12k}} +
\lambda \left[\sum_{k} \lambda^{\prime *}_{22k}
\lambda^{\prime}_{12k} - \sum_{k} \lambda^{\prime *}_{21k}
\lambda^{\prime}_{11k}
 \right] + O(\lambda^2)
\label{B.4}
\eey
The Cabibbo favored terms in (\ref{B.1})-(\ref{B.4}) have severe
constraints e.g. from study of $K^+ \to \pi^+ \nu \bar{\nu}$
decay \cite{18}:
\beq
\sum_k{\lambda^{\prime *}_{i1k} \lambda^\prime_{i^\prime2k}}
\leq 4.75 \times 10^{-5}
\left(\frac{m_{\tilde{q}}}{300GeV}\right)^2
\label{B.5}
\eeq
for $i \neq i^\prime$, and
\beq
\sum_k{\lambda^{\prime *}_{i1k} \lambda^\prime_{i2k}}
\leq 6.3 \times 10^{-5}
\left(\frac{m_{\tilde{q}}}{300GeV}\right)^2
\label{B.12}
\eeq
For $i = i^\prime$, bounds are about 30\% weaker because of
the impact of the SM and pure MSSM contributions \cite{18}.

It turns out that because of the
stringent bounds on the Cabibbo
favored terms, r.h.s.
of (\ref{B.1})-(\ref{B.4}) are dominated by the first
order Cabibbo suppressed terms.

The analysis for $\lambda_{ee}$ and $\lambda_{\mu \mu}$ is very
similar to that for $\lambda_{ss}$ and $\lambda_{dd}$. Assuming
that one of the couplings $\lambda_{12k}$ or $\lambda_{11k}$
dominates (say for k=3), one gets
\beq
- 0.91 \cdot 10^{-3}
\left(\frac{m_{\tilde{q}}}{300GeV}\right)^2 \leq
\lambda_{ee} \leq 3.83 \cdot 10^{-3}
\left(\frac{m_{\tilde{q}}}{300GeV}\right)^2 \label{B.6} \\
\eeq
In analogous way, assuming
that one of the couplings $\lambda_{22k}$ or $\lambda_{21k}$
dominates, one gets
\bey
\nn
&& - 0.0072
\left(\frac{m_{\tilde{q}}}{300GeV}\right)^2 \leq
\lambda_{\mu \mu} \leq 0.091
\left(\frac{m_{\tilde{q}}}{300GeV}\right)^2, \quad if \quad
m_{\tilde{q}} \leq 530GeV, \\
&& - 0.0072
\left(\frac{m_{\tilde{q}}}{300GeV}\right)^2 \leq
\lambda_{\mu \mu} \leq 0.29, \quad if \quad
m_{\tilde{q}} \geq 530GeV
\label{B.7}
\eey
The upper bound in the second line of (\ref{B.7}) comes
from the perturbativity bound on $\lambda^\prime_{22k}$ for
k=2,3 \cite{1}: $\lambda^\prime_{22k} \leq 1.12$.
We indicate the perturbativity bound saturation if only
it occurs for $m_{\tilde{q}} \leq 1TeV$.

The analysis for $\lambda_{\mu e}$ and $\lambda_{e \mu}$
is more subtle: instead of individual couplings squared
in absolute value,
the first order Cabibbo suppressed terms contain RPV
coupling pair products now. On our knowledge, there is
no bounds on pair products\footnote{One can meet some
bounds in the literature on
$\lambda^{\prime}_{1mk} \lambda^{\prime *}_{2mk}$
from study $\mu \to e \gamma$ decay
(see \cite{21} and references therein).
However, using our "rule of thumb", it is easy to see
that these are bounds on
$\widetilde{\lambda}^{\prime}_{12k}
\widetilde{\lambda}^{\prime *}_{22k}$, thus
they may not be used here.}
$\lambda^{\prime}_{12k} \lambda^{\prime *}_{22k}$ and
$\lambda^{\prime}_{11k} \lambda^{\prime *}_{21k}$.
Thus, we must use individual bounds on these four
couplings. As we deal with a pair product,
we may not anymore assume that only one RPV coupling
dominates. We must now allow for two RPV couplings
to be at their boundaries at a time. There is however
one subtlety: one may do this,
if only there is no correlations between the
constraints on $\lambda^\prime_{22k}$ and
$\lambda^\prime_{12k}$ or between those on
$\lambda^\prime_{21k}$ and $\lambda^\prime_{11k}$.

One can check that constraints on $\lambda^\prime_{22k}$ and
$\lambda^\prime_{12k}$ are indeed independent of each other
and constraints on $\lambda^\prime_{11k}$ are independent
of the values of $\lambda^\prime_{21k}$. The sources of
these constraints and references to the relevant
literature are given in \cite{1}. At first glance, the
situation with $\lambda^\prime_{21k}$ seems to be more
complicated: bounds on $\lambda^\prime_{21k}$ are derived from
$R_\pi \equiv \Gamma(\pi \to e \nu)/\Gamma(\pi \to
\mu \nu)$, assuming that
\cite{7}
\beq
|\lambda^\prime_{11k}|^2 \ll |\lambda^\prime_{21k}|^2
\label{B.8}
\eeq
On the other hand, one can see from Table~I in ref. \cite{1}
that
\beq
\max\left[|\lambda^\prime_{11k}|^2 \right]
\leq 0.13 \max\left[|\lambda^\prime_{21k}|^2\right]
\label{B.9}
\eeq
Thus, condition (\ref{B.8}) is satisfied to a good extent,
when $\lambda^\prime_{11k}$ and $\lambda^\prime_{21k}$ are
at their boundaries.

In what follows, one may use individual bounds on couplings
$\lambda^\prime_{11k}$, $\lambda^\prime_{21k}$,
$\lambda^\prime_{12k}$, $\lambda^\prime_{22k}$ presented in
ref. \cite{1}, to get constraints on the pair products
$\lambda^{\prime *}_{11k} \lambda^\prime_{21k}$ and
$\lambda^{\prime *}_{12k} \lambda^\prime_{22k}$. Using
these constraints and assuming that only one of these pairs
is non-zero (dominant) and only for a given $k$ (say k=3),
one gets
\bey
\nn
&& |\lambda_{\mu e}| \leq 0.019 \left(\frac{m_{\tilde{q}}}
{300GeV} \right)^2, \hspace{0.4cm}
|\lambda_{e \mu}| \leq 0.019 \left(\frac{m_{\tilde{q}}}
{300GeV} \right)^2,
\quad if \quad m_{\tilde{q}} \leq
530GeV \\
&& |\lambda_{\mu e}| \leq 0.033 \left(\frac{m_{\tilde{q}}}
{300GeV} \right), \hspace{0.5cm}
|\lambda_{e \mu}| \leq 0.033 \left(\frac{m_{\tilde{q}}}
{300GeV} \right),
\quad if \quad m_{\tilde{q}} \geq
530GeV
\label{B.10}
\eey
In deriving (\ref{B.10}), one must take into account that
products $\lambda^{\prime *}_{11k} \lambda^\prime_{21k}$ and
$\lambda^{\prime *}_{12k} \lambda^\prime_{22k}$ may be
both positive and negative.

Coincidence of bounds on $\lambda_{\mu e}$ and $\lambda_{e \mu}$
is not accidental: the first order Cabibbo suppressed terms in
equations (\ref{B.3}) and (\ref{B.4}) are complex conjugates of
each other. Thus,
$\lambda_{\mu e} \approx \lambda_{e \mu}^*$ or because we assume
that RPV coupling products relevant for our analysis are real,
one has
\beq
\lambda_{\mu e} \approx \lambda_{e \mu} \label{B.11}
\eeq

When deriving (\ref{B.10}) and (\ref{B.11}), we neglected
$O(\lambda^2)$ Cabibbo suppressed terms in the expressions
for $\lambda_{e \mu}$ and $\lambda_{\mu e}$.
If one assumes that
two RPV couplings dominate at a time, one should take into
account these terms as well. We leave for the
reader to verify that $O(\lambda^2)$ terms in the expressions
for $\lambda_{e \mu}$ and $\lambda_{\mu e}$ have at least
several times stronger bounds than
the first order Cabibbo suppressed
terms.


\end{document}